# Spontaneous Symmetry Breaking in Polar Fluids

Calum J. Gibb [1], Jordan L. Hobbs [2], Diana I. Nikolova[2], Tom J. Raistrick[2], Helen F. Gleeson[2], and Richard. J. Mandle* [1,2]

[1] School of Chemistry, University of Leeds, Leeds, UK, LS2 9JT
[2] School of Physics and Astronomy, University of Leeds, Leeds, UK, LS2 9JT

*r.mandle@leeds.ac.uk

**Abstract:** Spontaneous symmetry breaking and emergent polar order are each of fundamental importance to a range of scientific disciplines, as well as generating rich phase behaviour in liquid crystals (LCs). Here, we show the union of these phenomena to lead to two previously undiscovered polar liquid states of matter. Both phases have a lamellar structure with an inherent polar ordering of their constituent molecules. The first of these phases is characterised by polar order and a local tilted structure; the tilt direction precesses about a helix orthogonal to the layer normal, the period of which is such that we observe selective reflection of light. The second new phase type is anti-ferroelectric, with the constituent molecules aligning orthogonally to the layer normal. This has led us to term the phases the $SmC_P^H$ and SmA$_{AF}$ phases, respectively. Further to this, we obtain room temperature ferroelectric nematic (N$_F$) and $SmC_P^H$ phases *via* binary mixture formulation of the novel materials described here with a standard N$_F$ compound (DIO), with the resultant materials having melting points (and/or glass transitions) which are significantly below ambient temperature.

**One Sentence Summary:** We observe two new fluid phases of matter as a result from spontaneous symmetry breaking as a means to escape from polar order.

**Introduction**

Spontaneous symmetry breaking manifests in a wide range of scientific disciplines and ongoing problems for example: the Higgs mechanism in subatomic physics [1]; autocatalysis in chemistry [2]; homochirality in biology; [3] and the Dzyaloshinskii–Moriya interaction in ferromagnetism and ferroelectrics leading to the formation of magnetic skyrmions and other exotic spin textures [4]. A lesser-known subset of materials which spontaneously break symmetry are liquid crystals (LCs). LCs are fluids which can also exhibit spontaneous symmetry breaking through the formation of helical superstructures [5–10] or as a means to escape from bulk polar ordering [11]. LCs are synonymous with LCD technology, and, with this remaining the principal means of information display, the discovery of new liquid crystal phases at equilibrium is typically regarded as being of the highest significance. Current materials exploited in LCDs exhibit nematic (N) phases and are *apolar* (**Fig. 1a**), despite being formed from molecules with large electric dipole moments, due to the inversion symmetry of the director, and so the bulk material is polarisable but not polar.

Despite being the subject of theoretical interest in the 1920's [12], the lack of experimental discovery condemned polar nematic phases to obscurity for almost a century. The discovery of polar nematic phases at equilibrium in the late 2010's [13,14] has garnered significant excitement and has been described as having "promise to remake nematic science and technology" [15]. This is now referred to as the ferroelectric nematic (N$_F$) phase and is comprised of molecules with large electric dipole moments which align giving rise to a phase bulk polar order [16–19]. The lack of inversion symmetry means that the N_F phase possesses $C_{\infty v}$ symmetry with polarization along the director ($\hat{n}$) (i.e. largely parallel to the long molecular axis). Although in its infancy, the N_F phase has been suggested as candidate for multiple applications including as photo-variable capacitors [19], electrostatic actuators [20] as well as the next generation of display devices. Current materials (for example those in **Fig. 1c**) present the unwelcome combination of challenging working temperature ranges and often low chemical stability. This has driven a significant appetite for new materials that can sustain the polar mesophase at and below ambient temperatures [19,21].

Due to the low energy cost of elastic deformation of many liquid crystalline phases, polar order has recently been discovered in complex LC phase types [22–29]. Longitudinal ferroelectric variants of the apolar orthogonal smectic (lamellar) A (SmA) phase (**Fig.1d**) have recently been reported in both pure materials [30] and in binary mixtures [31]. Molecular rotation orthogonal to the long molecular axis in the ferroelectric SmA (SmA_F) phase is essentially frozen, this again resulting in a phase with no inversion symmetry ($C_{\infty v}$) and polarization parallel to $\hat{n}$, along the long molecular axis. Tilted smectic phases (such as SmC phases) can adopt a helical superstructure with polar order when the constituent molecules are chiral (e.g. SmC* [32], as the introduction of chirality reduces the symmetry from $C_h$ to $C_2$ and so leads to a polarisation perpendicular to the layer normal. If the helix is unwound, either through application of a field or surface treatment, a macroscopic polarisation (orders of magnitude smaller than in the N_F or SmA_F phases) results [15,33]. The polar SmC equivalent of the N_F/SmA_F phase has not yet been reported.

Herein we report a family of rod-like LC materials with large electric dipole moments roughly parallel to the long molecular axis. Some members display a polar SmC phase which spontaneously adopts a helical superstructure. Others display an antiferroelectric orthogonal SmA phase. Impressively, simple binary mixtures afford materials that are operable at (and far below) ambient temperatures.

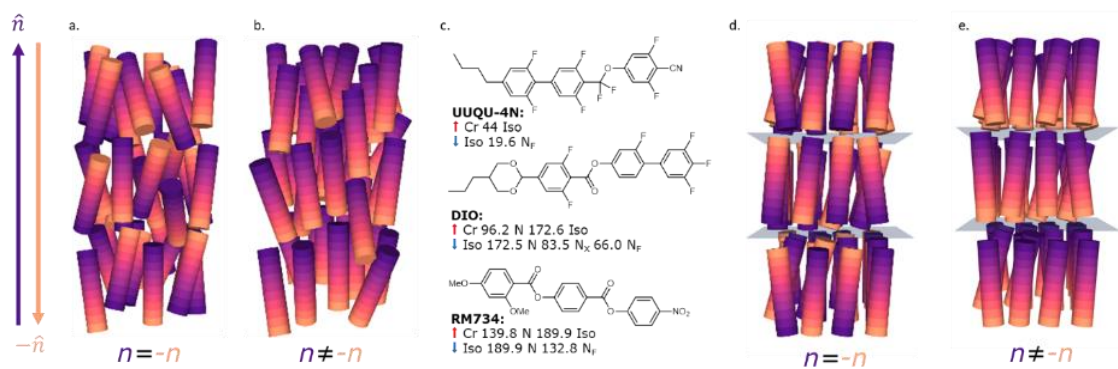

**Fig. 1:** Schematic depictions of (a) *apolar* and (b) *polar* nematic phases, (c) the structures of the exemplar N_F materials: UUQU-4-N [18]; DIO [14]; RM734 [34] and, (d) apolar and (e) polar smectic A phases. The lack of inversion symmetry within the polar variants of the N and SmA phases results in both phases possessing C_∞v symmetry with the polarization direction parallel to the director $\hat{n}$ along the long molecular axes.

**Results and Discussion**

During initial polarising microscopy studies, we observed compound **1** to exhibit selective reflection of light in a manner reminiscent of chiral nematic phases, although the chemical structure itself is achiral. This prompted the investigations detailed below, with our principal focus here on compounds **1, 4** and binary mixtures of **1** with DIO, however the phase behaviour of all 4 compounds synthesised are detailed in **Table 1**. Initial phase assignment for **1** was made on cooling from the isotropic liquid. First, a nematic (N) phase forms, easily identified by its characteristic *schlieren* texture when viewed between untreated glass slides (**Fig. 2a**). Further cooling of the N phase yields two orthogonal smectic phases: at higher temperatures a SmA phase which displays a focal-conic and fan texture, and a second smectic phase at lower temperatures in which the focal conics and fans become smooth and small regions appear close to the fan nucleation sites wherein the optical retardation changes rapidly across the sample (larger images shown in SI). Measurements of the spontaneous polarization ($P_S$) (**Fig. 2b**) reveal that the first smectic phase is apolar, confirming the assignment of a conventional SmA phase, but the lower temperature phase gives single peak in the current response indicating the phase is ferroelectric (**Fig. 2b**). Studies of the temperature dependence of the layer spacing (d($T$)) (**Fig. 2c**) confirm that the SmA$_F$ phase is indeed orthogonal as the layer spacing is essentially temperature independent and of the order of a single molecular length (≈ 3nm at the B3LYP-GD3BJ/aug-cc-pVTZ level of DFT). This supports the assignment of the SmA$_F$ phase below the conventional SmA phase for compound **1**.

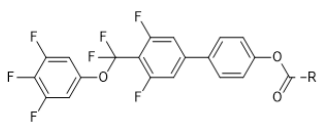

| Cmp. | R = | | Melt | SmC$_P^H$-SmA$_{F(AF)}$ | SmA$_{P2}$-SmA$_{F(AF)}$ | SmA$_{AF}$-SmA | SmA$_F$-SmA | SmA-N | N-Iso |
|---|---|---|---|---|---|---|---|---|---|
| 1 | C₃H₇ (dioxane-difluorophenyl) | T | 100.9 | [90.1] | - | - | 129.7 | 154.3 | 225.6 |
| | | ΔH | 31.1 | 0.01 | - | - | 0.6 | 0.1 | 1.0 |
| 2 | C₃H₇ (difluorophenyl) | T | 93.8 | - | [87.4] | [98.0] | - | 113.2 | 149.6 |
| | | ΔH | 30.6 | - | 0.01 | 0.8 | - | 0.8 | 0.7 |
| 3 | C₃H₇ (monofluorophenyl) | T | 101.7 | - | [98.9] | [105.4] | - | 141.0 | 167.6 |
| | | ΔH | 25.6 | - | 0.02 | 0.04 | - | 0.6 | 0.7 |
| 4 | C₃H₇ (phenyl) | T | 92.3 | [86.6] | - | [98.2] | - | 127.7 | 171.4 |
| | | ΔH | 21.7 | 0.08 | - | 0.04 | - | 0.3 | 0.9 |

**Table 1**: Transition Temperatures (T / °C) and associated enthalpies of transition (ΔH / kJ mol$^{-1}$) for compounds **1** – **4** determined by DSC at a heat/cool rate of 10 °C min$^{-1}$; phase assignments were made on the basis of polarized optical microscopy (POM), X-ray scattering, current-response and further experiments as described in the text.

Square brackets indicate a monotropic phase transition. The SmA$_{P2}$ phase is a 2$^{nd}$ polar orthogonal smectic phase of undetermined structure and is discussed further in the ESI.

Cooling the SmA$_F$ phase of **1** below ~88 °C yields a further phase transition wherein the smooth fan texture becomes significantly disrupted, yielding a texture with many small domains. The disruption quickly subsides allowing the texture to resolve into one similar to that seen in the SmA$_F$ phase, with some striations now visible along the backs of the smooth blocks similar to the SmA-SmC phase transition. When studied by DSC, the transitions between smectic phases have only a small associated enthalpy. X-ray measurements confirm the existence of a tilted phase as the layer spacing decreases monotonically as the molecules begin to tilt away from the layer normal (**Fig. 2c**). The growth in tilt is seemingly different to conventional SmC materials [35] [36] showing an almost linear temperature dependence, reaching a maximum of 23° at around 30 °C below the SmA-SmC transition, and is not saturated at the point which the material crystallises. We also obtained the tilt angle from the temperature dependence of optical birefringence measurements ($\Delta n$) (**Fig. 2d**), giving values slightly lower than that obtained from X-ray measurements the opposite of what is normally seen for tilted LC phases [37]. P$_S$ measurements confirm that polar order is retained upon entering the titled smectic phase however the peak in the current response splits into two non-equal peaks (**Fig. 2b**).

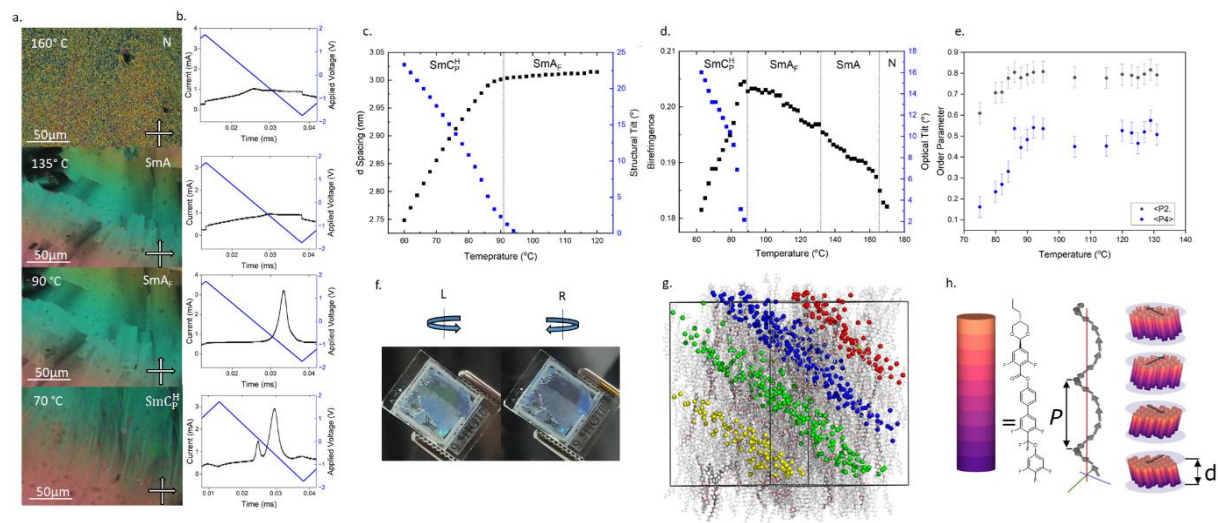

**Fig. 2:** Experimental data for **1**. (a) POM micrographs detailing characteristic textures of the N, SmA, SmA$_F$ and SmC$_P^H$ phases exhibited between glass coverslips, (b) Current response traces measured at 20 Hz and the temperature indicated in associated POM micrographs for the N, SmA, SmA$_F$ and SmC$_P^H$ phases, (c) Temperature dependence of the layer spacing (*d*) in the SmA$_F$ and SmC$_P^H$ phases [black]. Temperature dependence of molecular tilt calculated from the SAXS data in the SmC$_P^H$ phase is shown in blue, (d) Temperature dependence of optical birefringence ($\Delta n$) measured [black]. Temperature dependence of molecular tilt calculated from $\Delta n$ is shown in the blue trace, (e) Temperature dependence of the order parameters <P2> [black] and <P4> [blue] measured in a 10 µm cell treated for planar anchoring, (f) Photographs of **1** filled into a planar aligned cell taken with a circular polarisers of left and right polarisation demonstrating the handedness of the light reflected from the sample. The lack of uniformity across the cell is indicative of various domains of different helical rotation and pitch orientation, (g) Molecular dynamics simulations of compound **1**; instantaneous configuration at 249 ns of the polar SmC configuration at a temperature of 400 K. Layers are highlighted by rendering the oxygen atom of each CF$_2$O group as a sphere coloured by layer, and (h) The proposed model of the SmC$_P^H$ phase, where *P* is the helical periodicity (several hundreds of nanometres) and *d* is the smectic layer spacing (< 3 nm, from SAXS). Black arrows are used to illustrate the tilt orientation in four smectic blocks chosen as exemplars, each offset by 90° with a right-handed helical sense.

A simple visual inspection of **1** in the lowest temperature, tilted phase confined in a 5-micron cell under ambient lighting conditions clearly demonstrates its optical activity (**Fig. 2f**). Quantitative measurement of the wavelength dependence via spectroscopy proved challenging. The highly scattering texture of the tilted phase along with the presence of many small domains and the difficulty of aligning the periodicity responsible for the optical actively prevented quantitative measurements. However, placing a circular polariser of specific handedness between the sample and an observer showed changes in both the intensity and colour of the reflected light, indicating the presence of a chiral superstructure within the tilted phase. Due its ability to reflect visible light for at least some of the temperature regime of the tilted phase, we expect the periodicity of this structure to be of the order of a few hundreds of nanometres. To gain further insight into the nature of the order within **1,** we elected to measure the 2$^{nd}$ and 4$^{th}$ rank orientational order parameters as a function of temperature *via* polarised Raman spectroscopy (PRS) (**Fig. 2e**). In principle, PRS can discriminate between heliconical tilted phases and non-heliconical phases, although careful considerations must be made. For most heliconical structures the measured values of <P2> and <P4> are expected to decrease monotonically as the heliconical tilt angle grows [38,39], although for some systems a small tilt, coupled with an increase in the order parameter as the lower temperature phase is entered, could balance out such an effect. We find both <P2> and <P4> are approximately constant across the polar and non-polar SmA phases, taking values of ~0.8 and ~0.5, respectively. On entering the tilted phase there is a decrease in the measured values of <P2> and <P4>, which is consistent with the onset of a heliconical structure. The tilt angle can be inferred from the reduction of <P2> and <P4> on cooling (**Fig. S4**) suggesting an increasing tilt on cooling consistent with birefringence and X-ray measurements.

In parallel to physical observations, we performed MD simulations of **1** in a polar nematic configuration at a range of temperatures using the GAFF force field in Gromacs 2019.2 (see ESI for full details). Our aim was not to reproduce the periodic structure suggested by experiments to have a pitch of several hundred nanometres, versus the ~9 nm$^3$ volume simulated here, but rather to probe the molecular associations and local phase structure. Each simulation commences from a polar nematic starting configuration, however **1** readily adopts a lamellar structure (**Fig. 2g**) and so we observe a polar, tilted phase at temperatures up to 430 K, while at 440 K and 450 K we observe a polar SmA phase (**Fig. S4**). Based on the evidence presented thus far we propose that the lowest temperature phase be denoted as the polar heliconical smectic C ($\text{SmC}_\text{P}^\text{H}$) phase as the phase is a tilted smectic phase with a helical structure and its polar nature is clear from its response to an applied electric field. We suggest that the simplest structure for such a phase is that shown in **Fig. 2h.** Such a structure possesses C$_1$ symmetry with the direction of polarization parallel to the layer normal with the spontaneous formation of a helical superstructure, which spontaneously breaks symmetry, likely resulting from the need for the molecules to escape bulk polar order.

As part of our studies into **1** we synthesised a large number of structural analogues, some of which are reported in **Table 1**. Structure **4** shows a similar phase sequence as **1** (**Fig. 3a**), albeit at slightly decreased temperatures, however, **4** remarkably shows an antiferroelectric response in the second orthogonal SmA phase (**Fig. 3b+c**) which we therefore denote as SmA$_{\text{AF}}$. We are not aware of any prior observation of such a phase of matter. The simplest model of such a phase would be blocks of orthogonal smectic with opposing polar sense, in one sense a lamellar analogue of the SmZ$_\text{A}$ [31] (elsewhere referred to as N$_\text{X}$ or N$_\text{S}$) phase, as shown schematically in **Fig. 3d.** This structure is likely to experience some splay deformation to minimise free energy. Such a SmA$_{\text{AF}}$ phase type was recently suggested based on Onsager-Parsons-Lee local density functional theory [40], although it was also postulated to require possibly unphysical packing fractions of molecules based on steric effects alone. We conjecture that both the SmA$_{\text{AF}}$ and $\text{SmC}_\text{P}^\text{H}$ phases have their origins in spontaneous deformation as a means to escape polar order; whereas the tilted smectic phase can readily form a helix, the orthogonal SmA phase would instead form a 1D modulated structure. Many other complex structures can be envisaged however, and this will be the subject of a future publication. Additionally, the similarity of the modulation in the SmA$_{AF}$ phase to that in TGB phases [41] suggests a possible source of rich new phase behaviour *via* expulsion of twist (for example *via* chirality). Compounds **2** and **3**, which display further not previously disclosed novel polar phases, are discussed in the ESI.

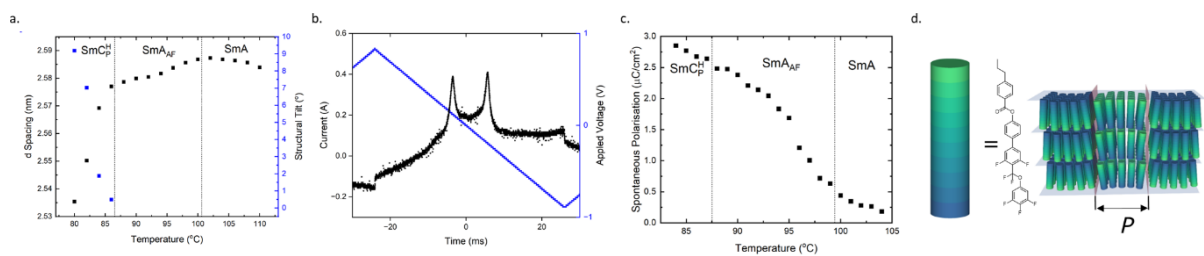

**Fig. 3**: (a) Temperature dependence of the layer spacing (*d*) for **4** in the SmA, SmA$_{AF}$ and SmC$_P^H$ phases [black]. Temperature dependence of molecular tilt calculated from the SAXS experiment in the SmC$_P^H$ phase is shown in the blue trace. (b) Current response of compound **4** at 96 °C in the SmA$_{AF}$ phase measured at 20 Hz, (c) Temperature dependence of the spontaneous polarization (P$_s$) measured at 20 Hz for **4** in the SmA, SmA$_{AF}$ and SmC$_P^H$ phases. The pretransitional P$_S$ measured in the SmA is induced polarisation due to the applied voltages analogous to the electroclinic effect at the SmA-SmC* transition. (d) The proposed model of the antiferroelectric orthogonal smectic A phase (SmA$_{AF}$), where *P* is the periodicity associated with the polar blocks.

We envisaged that binary mixtures of **1** with **DIO** would aid with phase identification through continuous miscibility. Despite this not being the case, the phase diagram (**Fig. 4a**) displays rich behaviour including polar materials whose meting/solidification points and/or vitrification are significantly below zero °C. Binary mixtures with concentrations of **DIO** greater than 50% exhibit N, N$_X$ and N$_F$ phases on cooling, each being identified by a combination of SAXS (**Fig. 4b**), P$_s$ (**Fig. 4c** and **Fig. S2**), and POM (**Fig. 4d**). Depending on composition, we next observe a transition into either the SmA$_F$ or SmC$_P^H$ phase. Notably, a mixture comprising 65% DIO provides the welcome observation of a stable N$_F$ phase at and below ambient temperatures. For increased concentrations of **1**, the polar nematic phases are not observed and either a SmA or N-SmA$_F$ phase transition is observed. The SmA$_F$ phase then transitions into the SmC$_P^H$ phase which, like the N$_F$ phase in the high concentrations of **DIO**, is stable at and below ambient temperature with mixtures comprising 40% **DIO** exhibiting the most stable example of the phase. The observation of room-temperature N$_F$ and SmC$_P^H$ phases in simple two-component mixtures are particularly exciting as an enabling new materials platform that promises to greatly simplify future experiments and simulations on these remarkable new phases of matter.

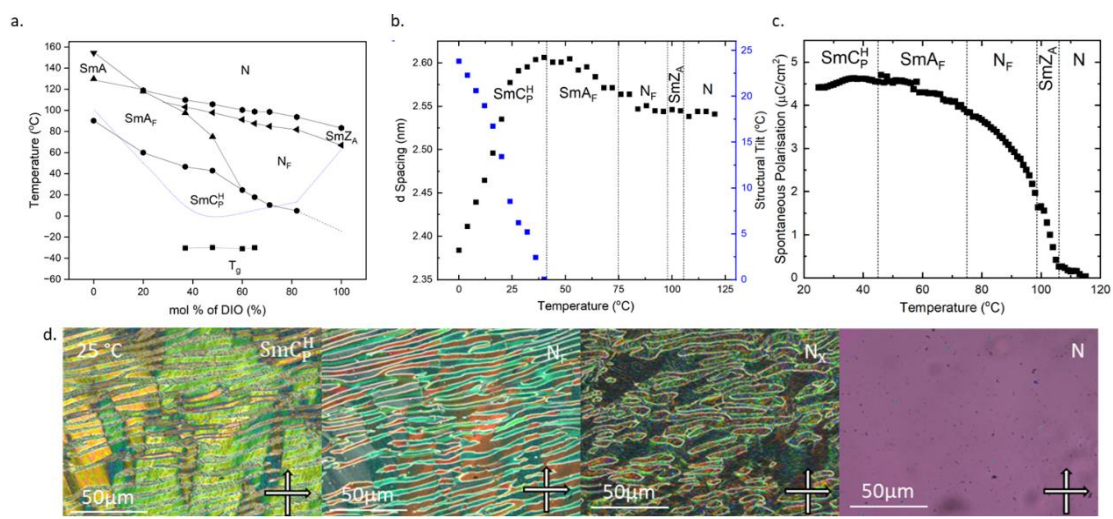

**Fig. 4**: (a) Phase diagram of binary mixtures between **1** and **DIO**. Melting points are denoted by the blue dotted line, and N-I transition temperatures have been omitted for clarity, (b) Temperature dependence of the layer spacing (*d*) for the N, $N_x$, $N_F$, $SmA_F$ and $SmC_P^H$ phases of a mixture of 50% **1** and **DIO**. For the N, $N_x$ and $N_F$ phase this does not correspond to a layer but corresponds to the long axis corelations present. (c) Temperature dependence of spontaneous polarization ($P_s$) for the same 50% mixture of **1** and **DIO** measured at 20 Hz, and (d) POM micrographs of a mixture of 35:65 molar ratio of **1** and **DIO** showing the N, $N_X$, $N_F$ and $SmC_P^H$ phases. The sample is confined within a thin cell treated for planar anchoring.

## Conclusions

We report the discovery of several new polar liquid crystalline phases which arise through a combination of polar ordering and/or spontaneous symmetry breaking. Firstly, we present the $SmC_P^H$ phase which exhibits selective reflection of light owing to its helical superstructure, as well as having a large spontaneous polarisation as a result of the near saturated polar order parameter. Through formulation of binary mixtures, we are able to obtain the $SmC_P^H$ phase at and below ambient temperature, greatly simplifying future experimental work. By varying the mixture composition, we also obtain ambient temperature $N_F$ materials, which are themselves a long-standing goal of the research community. Our observation of an antiferroelectric orthogonal smectic phase ($SmA_{AF}$), previously unknown to science, further demonstrates the rich new physics that arise from the combination of polar and orientational ordering in soft matter systems.

## Data availability

The data associated with this paper are openly available from the University of Leeds Data Repository at https://doi.org/10.5518/1488.

## Author Contribution Statement
C.J.G and J.L.H contributed equally to this work. C.J.G. and R.J.M performed chemical synthesis; J.L.H and C.J.G. performed mixture formulation studies, X-ray scattering experiments, microscopy, and DSC; J.L.H. and D.N. performed applied field studies; J.L.H performed birefringence measurements; T.R. performed polarised Raman spectroscopy; J.L.H and R.J.M. performed and evaluated electronic structure calculations; R.J.M. performed and analysed MD simulations; H.F.G and R.J.M secured funding. The manuscript was written, reviewed and edited with contributions from all authors.

## Competing interests
Authors declare that they have no competing interests.

## Acknowledgements

RJM thanks UKRI for funding *via* a Future Leaders Fellowship, grant number MR/W006391/1, and the University of Leeds for funding *via* a University Academic Fellowship. H.F.G acknowledges funding from EPSRC, grant Number EP/V054724/1. The SAXS/WAXS system used in this work was funded by EPSRC *via* grant number EP/X0348011. R.J.M. and H.F.G gratefully acknowledge support from Merck KGaA.

## Supplemental Materials
Materials and Methods
Supplementary Text
Figs S1 to S18

## References

[1]	P. W. Higgs, Broken Symmetries and the Masses of Gauge Bosons, Phys Rev Lett 13, 508 (1964).

[2]	Z. Shen, Y. Sang, T. Wang, J. Jiang, Y. Meng, Y. Jiang, K. Okuro, T. Aida, and M. Liu, Asymmetric Catalysis Mediated by a Mirror Symmetry-Broken Helical Nanoribbon, Nat Commun 10, 3976 (2019).

[3]	D. G. Blackmond, The Origin of Biological Homochirality, Cold Spring Harb Perspect Biol 11, 1 (2019).

[4]	H. J. Zhao, P. Chen, S. Prosandeev, S. Artyukhin, and L. Bellaiche, Dzyaloshinskii–Moriya-like Interaction in Ferroelectrics and Antiferroelectrics, Nat Mater 20, 341 (2021).

[5]	I. Dozov, On the Spontaneous Symmetry Breaking in the Mesophases of Achiral Banana-Shaped Molecules, Europhys Lett 56, 247 (2001).

[6]	M. Cestari et al., Phase Behavior and Properties of the Liquid-Crystal Dimer CB7CB: A Twist-Bend Nematic Liquid Crystal, Phys Rev E 84, 31704 (2011).

[7]	J. P. Abberley, R. Killah, R. Walker, J. M. D. Storey, C. T. Imrie, M. Salamończyk, C. Zhu, E. Gorecka, and D. Pociecha, Helicollical Smectic Phases Formed by Achiral Molecules, Nat Commun 9, 228 (2018).

[8]	S. P. Sreenilayam, Y. P. Panarin, J. K. Vij, V. P. Panov, A. Lehmann, M. Poppe, M. Prehm, and C. Tschierske, Spontaneous Helix Formation in Non-Chiral Bent-Core Liquid Crystals with Fast Linear Electro-Optic Effect, Nat Commun 7, 11369 (2016).

[9]	D. Pociecha, N. Vaupotič, M. Majewska, E. Cruickshank, R. Walker, J. M. D. Storey, C. T. Imrie, C. Wang, and E. Gorecka, Photonic Bandgap in Achiral Liquid Crystals—A Twist on a Twist, Advanced Materials 33, 2103288 (2021).

[10]	J. Karcz, J. Herman, N. Rychłowicz, P. Kula, E. Górecka, J. Szydlowska, P. W. Majewski, and D. Pociecha, Spontaneous Polar and Chiral Symmetry Breaking in Ordered Fluids -- Helicollical Ferroelectric Nematic Phases, (2023).

[11]	R. A. Reddy and C. Tschierske, Bent-Core Liquid Crystals: Polar Order, Superstructural Chirality and Spontaneous Desymmetrisation in Soft Matter Systems, J Mater Chem 16, 907 (2006).

[12]	M. Born, Ueber Anisotrope Flüssigkeiten: Versuch Einer Theorie Der Flüssigen Kristalle Und Des Elektrischen Kerr-Effekts in Flüssigkeiten, Sitzungsber Preuss Akad Wiss 30, 614 (1916).

[13]	R. J. Mandle, S. J. Cowling, and J. W. Goodby, Rational Design of Rod-Like Liquid Crystals Exhibiting Two Nematic Phases, Chemistry - A European Journal 23, 14554 (2017).

[14]	H. Nishikawa, K. Shiroshita, H. Higuchi, Y. Okumura, Y. Haseba, S. I. Yamamoto, K. Sago, and H. Kikuchi, A Fluid Liquid-Crystal Material with Highly Polar Order, Advanced Materials 29, 1702354 (2017).

[15]	O. D. Lavrentovich, Ferroelectric Nematic Liquid Crystal, a Century in Waiting, PNAS 117, 14629 (2020).

[16]	N. Sebastián, L. Cmok, R. J. Mandle, M. R. De La Fuente, I. Drevenšek Olenik, M. Čopič, and A. Mertelj, Ferroelectric-Ferroelastic Phase Transition in a Nematic Liquid Crystal, Phys Rev Lett 124, 037801 (2020).

[17]  X. Chen et al., First-Principles Experimental Demonstration of Ferroelectricity in a Thermotropic Nematic Liquid Crystal: Polar Domains and Striking Electro-Optics Contributed New Reagents/Analytic Tools; X, PNAS 117, 14021 (2002).

[18]  A. Manabe, M. Bremer, and M. Kraska, Ferroelectric Nematic Phase at and below Room Temperature, Liq Cryst 48, 1079 (2021).

[19]  J. Li, H. Nishikawa, J. Kougo, J. Zhou, S. Dai, W. Tang, X. Zhao, Y. Hisai, M. Huang, and S. Aya, Development of Ferroelectric Nematic Fluids with Giant-$\square$ Dielectricity and Nonlinear Optical Properties, Sci. Adv 7, 5047 (2021).

[20]  S. Nishimura, S. Masuyama, G. Shimizu, C. Chen, T. Ichibayashi, and J. Watanabe, Lowering of Electrostatic Actuator Driving Voltage and Increasing Generated Force Using Spontaneous Polarization of Ferroelectric Nematic Liquid Crystals, Advanced Physics Research 1, 2200017 (2022).

[21]  R. J. Mandle, A New Order of Liquids: Polar Order in Nematic Liquid Crystals, Soft Matter 18, 5014 (2022).

[22]  D. R. Link, G. Natale, R. Shao, J. E. Maclennan, N. A. Clark, E. Körblova, and D. M. Walba, Spontaneous Formation of Macroscopic Chiral Domains in a Fluid Smectic Phase of Achiral Molecules, Science (1979) 278, 1924 (1997).

[23]  L. E. Hough et al., Helical Nanofilament Phases, Science (1979) 325, 456 (2009).

[24]  R. A. Reddy et al., Spontaneous Ferroelectric Order in a Bent-Core Smectic Liquid Crystal of Fluid Orthorhombic Layers, Science (1979) 332, 72 (2011).

[25]  L. A. Madsen, T. J. Dingemans, M. Nakata, and E. T. Samulski, Thermotropic Biaxial Nematic Liquid Crystals, Phys Rev Lett 92, 145505 (2004).

[26]  V. Prasad, S.-W. Kang, K. A. Suresh, L. Joshi, Q. Wang, and S. Kumar, Thermotropic Uniaxial and Biaxial Nematic and Smectic Phases in Bent-Core Mesogens, J Am Chem Soc 127, 17224 (2005).

[27]  P. D. Duncan, M. Dennison, A. J. Masters, and M. R. Wilson, Theory and Computer Simulation for the Cubatic Phase of Cut Spheres, Phys Rev E 79, 31702 (2009).

[28]  R. Kotni, A. Grau-Carbonell, M. Chiappini, M. Dijkstra, and A. van Blaaderen, Splay-Bend Nematic Phases of Bent Colloidal Silica Rods Induced by Polydispersity, Nat Commun 13, 7264 (2022).

[29]  C. Meyer, C. Blanc, G. R. Luckhurst, P. Davidson, and I. Dozov, Biaxiality-Driven Twist-Bend to Splay-Bend Nematic Phase Transition Induced by an Electric Field, Sci Adv 6, eabb8212 (2024).

[30]  H. Kikuchi, H. Matsukizono, K. Iwamatsu, S. Endo, S. Anan, and Y. Okumura, Fluid Layered Ferroelectrics with Global C∞v Symmetry, Advanced Science 9, (2022).

[31]  X. Chen et al., The Smectic ZA Phase: Antiferroelectric Smectic Order as a Prelude to the Ferroelectric Nematic, PNAS 120, (2023).

[32]  J. P. F. Lagerwall and F. Giesselmann, Current Topics in Smectic Liquid Crystal Research, ChemPhysChem 7, 20 (2006).

[33]    X. Chen, E. Korblova, M. A. Glaser, J. E. Maclennan, D. M. Walba, and N. A. Clark, Polar In-Plane Surface Orientation of a Ferroelectric Nematic Liquid Crystal: Polar Monodomains and Twisted State Electro-Optics, PNAS 118, e2104092118 (2021).

[34]    R. J. Mandle, S. J. Cowling, and J. W. Goodby, A Nematic to Nematic Transformation Exhibited by a Rod-like Liquid Crystal, Physical Chemistry Chemical Physics 19, 11429 (2017).

[35]    N. Kapernaum, C. S. Hartley, J. C. Roberts, F. Schoerg, D. Krueerke, R. P. Lemieux, and F. Giesselmann, Systematic Variation of Length Ratio and the Formation of Smectic A and Smectic C Phases, ChemPhysChem 11, 2099 (2010).

[36]    S. J. Cowling, A. W. Hall, J. W. Goodby, Y. Wang, and H. F. Gleeson, Examination of the Interlayer Strength of Smectic Liquid Crystals through the Study of Partially Fluorinated and Branched Fluorinated End-Groups, J Mater Chem 16, 2181 (2006).

[37]    J. T. Mills, H. F. Gleeson, J. W. Goodby, M. Hird, A. Seed, and P. Styring, A Comparison of the Optical and Steric Tilt in Antiferroelectric Liquid Crystals, Molecular Crystals and Liquid Crystals Science and Technology. Section A. Molecular Crystals and Liquid Crystals 330, 449 (1999).

[38]    Z. Zhang, V. P. Panov, M. Nagaraj, R. J. Mandle, J. W. Goodby, G. R. Luckhurst, J. C. Jones, and H. F. Gleeson, Raman Scattering Studies of Order Parameters in Liquid Crystalline Dimers Exhibiting the Nematic and Twist-Bend Nematic Phases, J Mater Chem C Mater 3, 10007 (2015).

[39]    G. Singh, J. Fu, D. M. Agra-Kooijman, J.-K. Song, M. R. Vengatesan, M. Srinivasarao, M. R. Fisch, and S. Kumar, X-Ray and Raman Scattering Study of Orientational Order in Nematic and Heliconical Nematic Liquid Crystals, Phys Rev E 94, 60701 (2016).

[40]    P. Kubala, M. Cieśla, and L. Longa, Splay-Induced Order in Systems of Hard Tapers, Phys Rev E 108, 54701 (2023).

[41]    J. W. Goodby, M. A. Waugh, S. M. Stein, E. Chin, R. Pindak, and J. S. Patel, Characterization of a New Helical Smectic Liquid Crystal, Nature 337, 449 (1989).

[42]    Martinot-Lagarde Ph., Direct Electrical Measurement of the Permanent Polarization of a Ferroelectric Chiral Smectic C Liquid Crystal, J. Physique Lett. 38, 17 (1977).

[43]    K. Miyasato, S. Abe, H. Takezoe, A. Fukuda, and E. Kuze, Direct Method with Triangular Waves for Measuring Spontaneous Polarization in Ferroelectric Liquid Crystals, Jpn J Appl Phys 22, L661 (1983).

[44]    C. Meyer, G. R. Luckhurst, and I. Dozov, The Temperature Dependence of the Heliconical Tilt Angle in the Twist-Bend Nematic Phase of the Odd Dimer CB7CB, J Mater Chem C Mater 3, 318 (2015).

[45]    M. J. Frisch, G. W. Trucks, H. B. Schlegel, G .E. Scuseria, M. A. Robb, J. R. Cheeseman, G. Scalmani, V. Barone, B. Mennucci, G. A. Petersson, H. Nakatsuji, M. Caricato, X. Li, H. P. Hratchian, A. F. Izmaylov, J. Bloino, G. Zheng, J. L. Sonnenberg, M. Hada, M. Ehara, K. Toyota, R. Fukuda, J. Hasegawa, M. Ishida, T. Nakajima, Y. Honda, O. Kitao, H. Nakai, T. Vreven, J. A. Montgomery Jr., J. E. Peralta, F. Ogliaro, M. J. Bearpark, J. Heyd, E. N. Brothers, K. N. Kudin, V. N. Staroverov, R. Kobayashi, J. Normand, K. Raghavachari, A. P. Rendell, J. C. Burant, S. S. Iyengar, J. Tomasi, M. Cossi, N. Rega, N. J. Millam, M. Klene, J. E. Knox, J. B. Cross, V. Bakken, C. Adamo, J. Jaramillo, R. Gomperts, R. E. Stratmann, O. Yazyev, A. J. Austin, R. Cammi, C. Pomelli, J. W. Ochterski, R. L. Martin,

K. Morokuma, V. G. Zakrzewski, G. A. Voth, P. Salvador, J. J. Dannenberg, S. Dapprich, A. D. Daniels, Ö. Farkas, J. B. Foresman ,J. V. Ortiz, J. Cioslowskiand, D. J. Fox, Gaussian 016, Revision E.01, Gaussian, Inc., Wallingford CT, 2016.

[46]    S. Wang, J. Witek, G. A. Landrum, and S. Riniker, Improving Conformer Generation for Small Rings and Macrocycles Based on Distance Geometry and Experimental Torsional-Angle Preferences, J Chem Inf Model 60, 2044 (2020).

[47]    J. Wang, R. M. Wolf, J. W. Caldwell, P. A. Kollman, and D. A. Case, Development and Testing of a General Amber Force Field, J Comput Chem 25, 1157 (2004).

[48]    C. I. Bayly, P. Cieplak, W. D. Cornell, and P. A. Kollman, A Well-Behaved Electrostatic Potential Based Method Using Charge Restraints for Deriving Atomic Charges: The RESP Model, J. Phys. Chem 97, 10269 (1993).

[49]    C. Lee, eitao Yang, and R. G. Parr, Development of the Colic-Salvetti Correlation-Energy Formula into a Functional of the Electron Density, Phys Rev B 37, 15 (1988).

[50]    A. D. Becke, Density-Functional Thermochemistry. III. The Role of Exact Exchange, Journal of Chemical Physics 98, 5648 (1993).

[51]    J. Wang, W. Wang, P. A. Kollman, and D. A. Case, Automatic Atom Type and Bond Type Perception in Molecular Mechanical Calculations, J Mol Graph Model 25, 247 (2006).

[52]    D. A. Case, T. E. Cheatham, T. Darden, H. Gohlke, R. Luo, K. M. Merz, A. Onufriev, C. Simmerling, B. Wang, and R. J. Woods, The Amber Biomolecular Simulation Programs, J Comput Chem 26, 1668 (2005).

[53]    D. Silva and B. F. Vranken, ACPYPE-AnteChamber PYthon Parser InterfacE, Research Notes 5, 367 (2012).

[54]    B. Hess, H. Bekker, H. J. C. Berendsen, and J. G. E. M. Fraaije, LINCS: A Linear Constraint Solver for Molecular Simulations, J Comput Chem 18, 1463 (1997).

[55]    S. Nosé and M. L. Klein, Constant Pressure Molecular Dynamics for Molecular Systems, Mol Phys 50, 1055 (1983).

[56]    M. Parrinello and A. Rahman, Polymorphic Transitions in Single Crystals: A New Molecular Dynamics Method, J Appl Phys 52, 7182 (1981).

[57]    W. G. Hoover, Canonical Dynamics: Equilibrium Phase-Space Distributions, Phys Rev A (Coll Park) 31, 1695 (1985).

[58]    S. Nosé, A Molecular Dynamics Method for Simulations in the Canonical Ensemble, Mol Phys 52, 255 (1984).

[59]    R. T. McGibbon, K. A. Beauchamp, M. P. Harrigan, C. Klein, J. M. Swails, C. X. Hernández, C. R. Schwantes, L. P. Wang, T. J. Lane, and V. S. Pande, MDTraj: A Modern Open Library for the Analysis of Molecular Dynamics Trajectories, Biophys J 109, 1528 (2015).

[60]    R. J. Mandle, Implementation of a Cylindrical Distribution Function for the Analysis of Anisotropic Molecular Dynamics Simulations, PLoS One 17, (2022).

[61]    R. J. Mandle, N. Sebastián, J. Martinez-Perdiguero, and A. Mertelj, On the Molecular Origins of the Ferroelectric Splay Nematic Phase, Nat Commun 12, 4962 (2021).

[62]    X. Chen et al., Observation of a Uniaxial Ferroelectric Smectic A Phase, Proceedings of the National Academy of Sciences 119, (2022).

[63]  N. Hayashi, A. Kocot, M. J. Linehan, A. Fukuda, J. K. Vij, G. Heppke, J. Naciri, S. Kawada, and S. Kondoh, Experimental Demonstration, Using Polarized Raman and Infrared Spectroscopy, That Both Conventional and de Vries Smectic-$A$ Phases May Exist in Smectic Liquid Crystals with a First-Order $A\text{\ensuremath{-}}{C}^{*}$ Transition, Phys Rev E 74, 51706 (2006).

[64]  T. Raistrick, Z. Zhang, D. Mistry, J. Mattsson, and H. F. Gleeson, Understanding the Physics of the Auxetic Response in a Liquid Crystal Elastomer, Phys Rev Res 3, 23191 (2021).

[65]  A. Sanchez-Castillo, M. A. Osipov, S. Jagiella, Z. H. Nguyen, M. Kašpar, V. Hamplovă, J. Maclennan, and F. Giesselmann, Orientational Order Parameters of a de Vries--Type Ferroelectric Liquid Crystal Obtained by Polarized Raman Spectroscopy and x-Ray Diffraction, Phys Rev E 85, 61703 (2012).

[66]  N. Hayashi, A. Kocot, M. J. Linehan, A. Fukuda, J. K. Vij, G. Heppke, J. Naciri, S. Kawada, and S. Kondoh, Experimental Demonstration, Using Polarized Raman and Infrared Spectroscopy, That Both Conventional and de Vries Smectic-$A$ Phases May Exist in Smectic Liquid Crystals with a First-Order $A\text{\ensuremath{-}}{C}^{*}$ Transition, Phys Rev E 74, 51706 (2006).

[67]  N. Hayashi and T. Kato, Investigations of Orientational Order for an Antiferroelectric Liquid Crystal by Polarized Raman Scattering Measurements, Phys Rev E 63, 21706 (2001).

[68]  A. Kocot et al., Observation of the de Vries Behavior in SmA* Phase of a Liquid Crystal Using Polarised Raman Scattering and Infrared Spectroscopy, J Chem Phys 147, 094903 (2017).

[69]  C. A. Parton-Barr, H. F. Gleeson, and R. J. Mandle, Room-Temperature Ferroelectric Nematic Liquid Crystal Showing a Large and Diverging Density, Soft Matter 20, 672 (2024).

# Spontaneous Symmetry Breaking in Polar Fluids

# Supplemental Information

Calum J. Gibb [1], Jordan L. Hobbs [2], Diana I. Nikolova[2], Tom J. Raistrick[2], Helen F. Gleeson[2], and Richard. J. Mandle* [1,2]

[1] School of Chemistry, University of Leeds, Leeds, UK, LS2 9JT
[2] School of Physics and Astronomy, University of Leeds, Leeds, UK, LS2 9JT
*r.mandle@leeds.ac.uk

**Contents**

## 1 Experimental methods

### 1.1. Chemical Synthesis

Chemicals were purchased from commercial suppliers (Fluorochem, Merck, ChemScene, Ambeed) and used as received. Solvents were purchased from Merck and used without further purification. Reactions were performed in standard laboratory glassware at ambient temperature and atmosphere and were monitored by TLC with an appropriate eluent and visualised with 254 nm light. Chromatographic purification was performed using a Combiflash NextGen 300+ System (Teledyne Isco) with a silica gel stationary phase and a hexane/ethyl acetate gradient as the mobile phase, with detection made in the 200-800 nm range. Chromatographed materials were subjected to re-crystallisation from an appropriate solvent system.

### 1.2. Chemical Characterisation Methods

The structures of intermediates and final products were determined using $^1$H, $^{13}$C{$^1$H}, and $^{19}$F NMR spectroscopy. For chemical intermediates, NMR spectroscopy was performed using a Bruker Avance III HDNMR spectrometer operating at 400 MHz, 100.5 MHz or 376.4 MHz ($^1$H, $^{13}$C{$^1$H} and $^{19}$F, respectively). For final products (**1-4**), NMR spectroscopy was performed using a Bruker AV4 NEO 11.75T spectrometer operating at 500 MHz, 125.5 MHz or 470.5 MHz ($^1$H, $^{13}$C{$^1$H} and $^{19}$F, respectively).

### 1.3. Mesophase Characterisation

Phase transition temperatures and associated enthalpies of transition for compounds **1-4** were determined by differential scanning calorimetry (DSC) using a TA instruments Q2000 heat flux calorimeter with a liquid nitrogen cooling system for temperature control. Samples were measured with 10 °C min$^{-1}$ heating and cooling rates. The transition temperatures and enthalpy values reported are averages obtained for duplicate runs. Phase transition temperatures were measured on cooling cycles for consistency between monotropic and enantiotropic phase transitions, while crystal melts were obtained on heating. Phase identification by polarised optical microscopy (POM) was performed using a Leica DM 2700 P polarised optical microscope equipped with a Linkam TMS 92 heating stage. Samples were studied sandwiched between two untreated glass coverslips.

### 1.4. X-ray Scattering

X-ray scattering measurements, both small angle (SAXS) and wide angle (WAXS) where recorded using an Anton Paar SAXSpoint 5.0 beamline machine. This was equipped with a primux 100 Cu X-ray source with a 2D EIGER2 R detector. The X-rays had a wavelength of 0.154 nm. Samples were filled into either thin-walled quartz capillaries or held between Kapton tape. Temperature was controlled using an Anton Paar heated sampler with a range of -10 °C to 120°C and the samples held in a chamber with an atmospheric pressure of <1 mBa. Samples were held at 120°C to allow for temperature equilibration across the sample and then slowly cooled while stopping to record the scattering patterns. Compounds **2**, **3** and **4** were measured using a beam-stop while **1** was measured beam-stopless to allow for the whole pattern to be observed. Background scattering off the sample holders was subtracted from the obtained patterns after being appropriately scaled.

The samples were not formally aligned and so these measurements can be considered as "powder" samples. It should be noted that some spontaneous alignment of the LCs both within the capillaries and between the Kapton tape did occur leading to the classic "lobe" pattern seen in the 2D patterns. 1D patterns were obtained by radially integrating the 2D SAXS patterns. Peak position and FWHM was recorded and then converted into d spacing following Bragg's law. In the tilted smectic phase, the tilt was obtained from:

$$\frac{d_c}{d_A} = \cos\theta \qquad (1)$$

where $d_c$ is the layer spacing in the tilted smectic phase, $d_A$ is the extrapolated spacing from the non-tilted preceding smectic phase, extrapolated to account for the weak temperature dependence of the preceding phases due to shifts in conformation and order, and $\theta$ the structural tilt angle.

### 1.5. Measurement of Spontaneous Polarization ($P_S$)

Spontaneous polarisation measurements are undertaken using the current reversal technique [42,43]. Triangular waveform AC voltages are applied to the sample cells with an Agilent 33220A signal generator (Keysight Technologies), and the resulting current outflow is passed through a current-to-voltage amplifier and recorded on a RIGOL DHO4204 high-resolution oscilloscope (Telonic Instruments Ltd, UK). Heating and cooling of the samples during these measurements is achieved with an Instec HCS402 hot stage controlled to 10 mK stability by an Instec mK1000 temperature controller. The LC samples are held in 4µm thick cells with no alignment layer, supplied by Instec. The measurements consist of cooling the sample at a rate of 1 Kmin$^{-1}$ and applying a set voltage at a frequency of 20 Hz to the sample every 1 K. The voltage was set such that it would saturate the measured $P_S$ and was determined before final data collection.

There are three contributions to the measured current trace: accumulation of charge in the cell ($I_c$), ion flow ($I_i$), and the current flow due to polarisation reversal ($I_p$). To obtain a $P_S$ value, we extract the latter, which manifests as one or multiple peaks in the current flow, and integrate as:

$$P_S = \int \frac{I_p}{2A} dt \qquad (2)$$

where A is the active electrode area of the sample cell. For the N, SmZ$_A$ and, to a lesser extent, the N$_F$ phase, significant amounts of ion flow is present. For materials and mixtures that showed a paraelectric N phase followed by the anti-ferroelectric SmZ$_A$ phase, the N phase always showed some pre-transitional polarisation as well as the significant ion flow mentioned previously. Since the following phase was anti-ferroelectric, this pre-transitional polarisation was anti-ferroelectric in character and so was decoupled from the ion flow in the same way as the SmZ$_A$ phase and as such the $P_S$ of the N and SmZ$_A$ phases was obtained by integrating the peak least affected by ion flow and then doubled to get the total area under both peaks [31].

For the various polar smectic phases found in these materials generally we observed low charge accumulation and ion flow allowing for the baseline to be easily defined and the integrated area of the peak or peaks to be obtained accurately. However, much like for the N phase preceding the SmZ$_A$ phase, the paraelectric SmA phases showed significant pre-transitional polarisation which was measured in the same way as the various polar smectic phases. The polarisation character of the pre-transitional polarisation was always the same as the polar phase that followed.

### 1.6. Polarised Raman Spectroscopy (PRS)

Order parameters were determined via polarized Raman spectroscopy (PRS) as described elsewhere [38,39]. Measurements were performed on 10μm Instec cells with anti-parallel planar surface alignment. Raman spectrometry was performed using a Renishaw invia Raman spectrometer equipped with a 20 mW 532 nm laser and an optical microscope with a 10x objective. Measurements were performed in well-aligned regions of the sample devoid of defects using an exposure time of 3x 30s. The Raman spectra are recorded at angles of the nematic director from 0° to 180° with respect to the incident laser polarization for parallel and perpendicular polarized backscattered light. The 1606cm$^{-1}$ peak was determined to correspond to the C-C breathing mode of the non-fluorinated phenyl ring of **1** via DFT calculations (detailed in section 1.8 and 2.4). The peak was used to determine order parameters by fitting the intensity of the parallel ($I_\parallel$) and perpendicular ($I_\perp$) components to the following equations [38]:

$$I_\parallel(\theta) \propto \frac{1}{5} + \frac{4r}{14} + \frac{8r^2}{15} + \langle P_2 \rangle \left[\frac{1}{21}(3 + r - 4r^2)(1 + 3\cos(2\theta))\right] + \langle P_4 \rangle \left[\frac{1}{280}(1-r)^2(9 + 20\cos(2\theta) + 35\cos(4\theta))\right] \quad (3)$$

$$I_\perp(\theta) \propto \frac{1}{15}(1-r)^2 + \langle P_2 \rangle \left[\frac{1}{21}(1-r)^2\right] + \langle P_4 \rangle \left[\frac{1}{280}(1-r)^2(3 - 35\cos(4\theta))\right] \quad (4)$$

Where $r$ is the differential molecular polarizability ratio, $\langle\,\rangle$ denotes an ensemble average, and $P_n$ is the $n_{\text{th}}$ Legendre polynomial function. Generally, fits are performed on the depolarisation ratio, $R(\theta)$, to reduce the dependence of the fitting on the incident laser intensity:

$$R(\theta) = \frac{I_\perp(\theta)}{I_\parallel(\theta)} \quad (5)$$

### 1.7. Birefringence Measurements

Birefringence was measured using a Berek compensator mounted in a Leica DM 2700 P polarised optical microscope. The LC sample of compound **1** was measured in a 10 μm anti-parallel rubbed planar cells purchased from Instec. Alignment quality was good in all the phases preceding the $SmC_P^H$ phase evidenced by the high level of extension in the positions where the rubbing direction of the filled cell was parallel and perpendicular to the polariser axis. While alignment quality decreased in the $SmC_P^H$ phase an acceptable dark start was obtained parallel and perpendicular to the polariser axis. With the Berek inserted this change in alignment quality translated to a broadening of the extinction fringes but a measurement of optical retardance, and thus birefringence, was still obtained albeit with slightly larger values of associated error.

Birefringence was converted to optical tilt using the equation [44]:

$$\Delta n_{CHP} = \Delta n_{A(F)}(3\cos^2\theta - 1)/2 \quad (6)$$

where $\Delta n_{CHP}$ corresponds to the birefringence of the $SmC_P^H$ phase, $\Delta n_{A(F)}$ the extrapolated birefringence of the SmA$_F$ phase and θ the optical tilt.

### 1.8. Electronic Structure Calculations

Electronic structure calculations were performed using Gaussian G16 revision C.01 [45]. For each input structure we first generated a set of unique low energy conformers using the ETKDGv3 rules-based method [46]; each conformer then underwent geometry optimisation at the B3LYP-GD3BJ/aug-cc-pVTZ level of DFT followed by a frequency calculation to confirm the geometry to be at a minimum. For each molecule we then obtain properties (dipole tensor, polarisability tensor) as the probability weighted average. Use of a single minimum energy conformer typically overestimates the molecular dipole moment (e.g. 11.9 D for **1**) and anisotropic polarisability (55.8 Å$^3$ for **1**)

### 1.9. MD Simulation Setup and Analysis

Fully atomistic molecular dynamics (MD) simulations were performed in Gromacs 2019, with parameters modelled using the General Amber Force Field (GAFF). [47] Atomic charges were determined using the RESP method [48] for geometry optimised at the B3LYP/6-31G(d) level of DFT [49,50] using the Gaussian G16 revision c01 software package [45]. Topologies were generated using AmberTools 16 [51,52] and converted into Gromacs readable format with Acpype. [53]

We initially constructed a low-density lattice of 800 molecules of **1-4** with random positional and orientational order. Following energy minimization by the steepest decent method we performed short (5 ns) equilibration simulations in the NVE and NVT (T = 600 K) ensembles. We then performed a short 'compression' simulation (25 ns) at 600K with an isotropic barostat (P = 100 Bar) to compress the simulation to a liquid like density (~ 1.1 g cm$^3$). We then obtained a *polar* nematic configuration by applying a static electric field (1 V nm$^{-1}$) along the x-axis of the isotropic starting configuration for a total of 50 ns at 600 K and 1 Bar; this configuration was used as a starting point for subsequent simulations. Production MD simulations were performed without the biasing field and employed an anisotropic barostat (pressure of 1 Bar), at temperatures of 400 K (and for **1**, also in 10 K increments up to a maximum of 550 K) for a further 250 ns, unless otherwise noted.

Simulations employed periodic boundary conditions in xyz. Bonds lengths were constrained to their equilibrium values with the LINCS algorithm [54]. During production MD simulations the system pressure was maintained at 1 Bar using an isotropic Parrinello-Rahamn barostat. [55,56]. Compressabilities in xyz dimensions were set to 4.5e-5, with the off-diagonal compressibilities were set to zero to ensure the simulation box remained rectangular. Simulation temperature was controlled with a Nosé–Hoover thermostat. [57,58] Long-range electrostatic interactions were calculated using the Particle Mesh Ewald method with a cut-off value of 1.2 nm. A van der Waals cut-off of 1.2 nm was used. MD trajectories were visualised using PyMOL 4.5. Q-tensor analysis was performed using MDTraj 1.9.8 [59]. Cylindrical distribution functions (CDF) were computed using the *cylindr* code. [60] Simulation densities, dipole moments and volumes were obtained with the *gmx energy* program, with dipole and volume being used to compute spontaneous polarisation.

We calculate the second-rank orientational order parameter <$P_2$> *via* the Q-tensor according to eq. **(7)**;

$$Q_{\alpha\beta} = \frac{1}{N}\sum_{m=1}^{N}\frac{3a_{m\alpha}a_{m\beta}-\delta_{\alpha\beta}}{2} \quad \textbf{(7)}$$

where N is the number of molecules, m is the molecule number within a given simulation, α and β represent the Cartesian x, y and z axes, delta is the Kronecker delta, *a* is a vector that describes the molecular long axis, which is computed for each monomer as the eigenvector associated with the smallest eigenvalue of the inertia tensor. The director at each frame was defined as the eigenvector associated with the largest eigenvalue of the ordering tensor. The order parameter $\langle P_2 \rangle$ corresponds to the largest eigenvalue of $Q_{\alpha\beta}$, and the biaxial order parameter $\langle B \rangle$ corresponds to the difference between the two smallest eigenvalues. The polar order parameter, $\langle P_1 \rangle$, was calculated as the total dipole moment of the simulation box over the sum of the individual molecular dipoles. The polarization, *P*, was calculated from the total dipole of the box over the volume. For smectic phases formed in MD simulations the layer spacing was determined by first fitting a planes to a set of reference atom for each molecule (the oxygen atom of the $CF_2O$ group); the fit with the highest $R^2$ indicates the number of layers per simulation (4 in all cases). The layer spacing was then taken as the average distance between planes in the final 200 ns of the simulation.

## 2 Additional data and supplemental results and discussion

### 2.1. Polarized Optical Microcopy Images

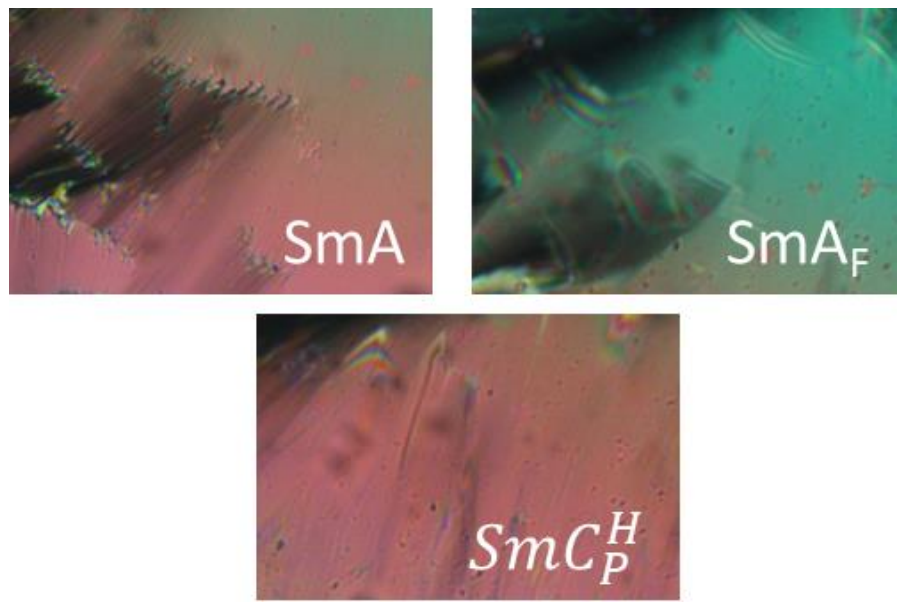

**Fig. S1**: POM images of **1** in the SmA, SmA$_F$ and $SmC_P^H$ phases showing the same region after cooling. The defects formed between neighbouring fans in the SmA form into regions where the optical retardence rapidly changes showing various birefringence colours. Also clear from these images is how the striated backs of the fans in the SmA texture smooth into uniform areas in the SmA$_F$ phase. These striations return in the $SmC_P^H$ phase and the regions of rapid colour change break into fragmented broken features.

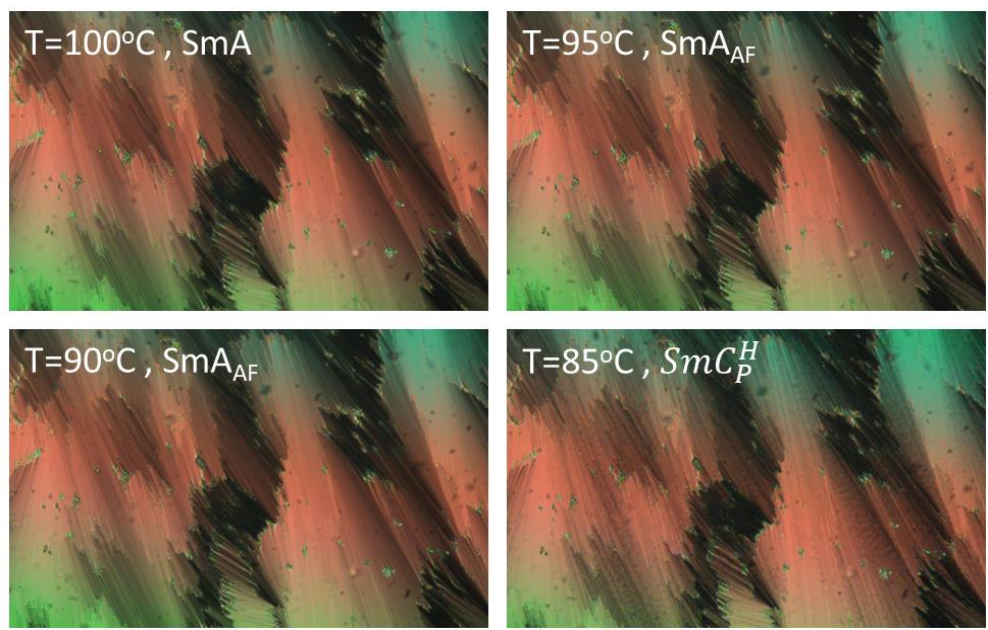

**Fig. S2**: POM images of **4** in the four smectic phase it exhibits. These images were taken between untreated coverslips on cooling at the indicated temperatures.

While the textural changes between the SmA and the SmA$_F$ can be quite significant (discussed in the main text), there are few changes between the texture of the SmA and SmA$_{AF}$ phase. It has been suggested that ferroelectric ordering inhibits the formation of the Dupin cyclides responsible for the fan and focal conic defects found in the natural texture of the SmA phase [30]. Seemingly the SmA$_{AF}$ phase does not have the same issue with forming fans and the natural texture is much more similar with only slight changes in the defects found where neighbouring fans meet. This could be due to paramorphotism of the fan structure found in the SmA phase, though attempts to form textures other than this failed for compound **4**. Upon further cooling the $\mathrm{SmC_P^H}$ phase is reached and the backs of the fans become increasing striated much like the transition from SmA to SmC* where the bulk change is molecular tilt and formation of a helix.

## 2.2. X-ray Scattering

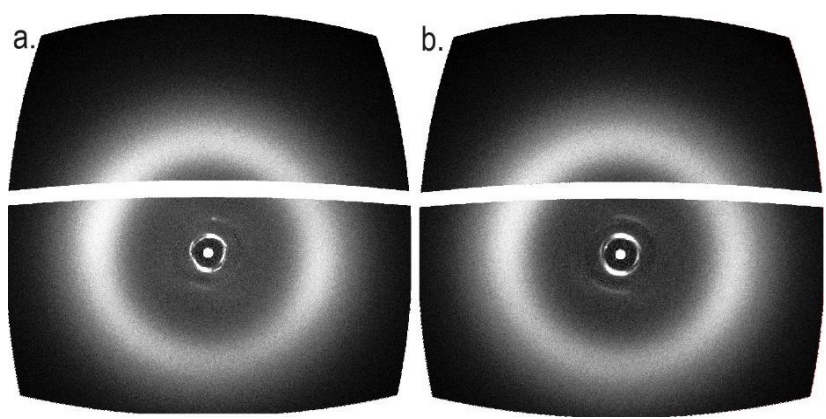

**Fig. S3:** X-ray scattering patterns of **1** in the (a) SmA$_F$ and (b) $\mathrm{SmC_P^H}$ phases.

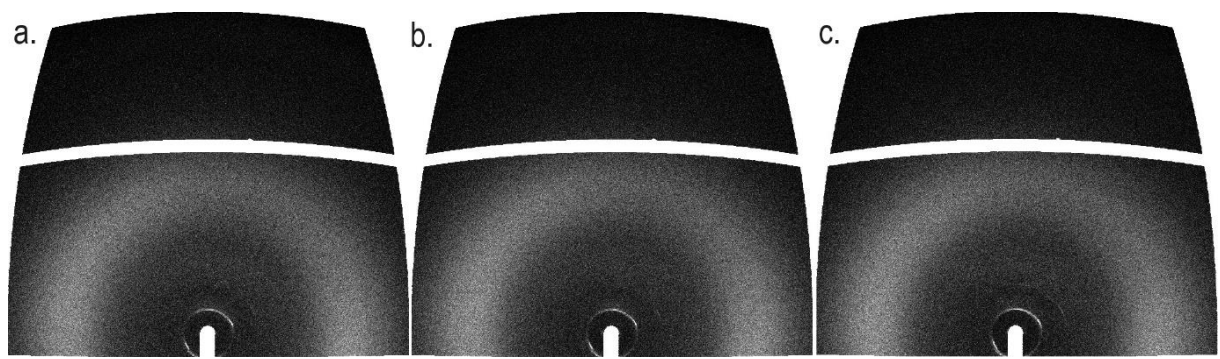

**Fig. S4:** X-ray scattering patterns of **2** in the (a) SmA, (b) SmA$_{AF}$ and (c) SmA$_{P2}$ phases.

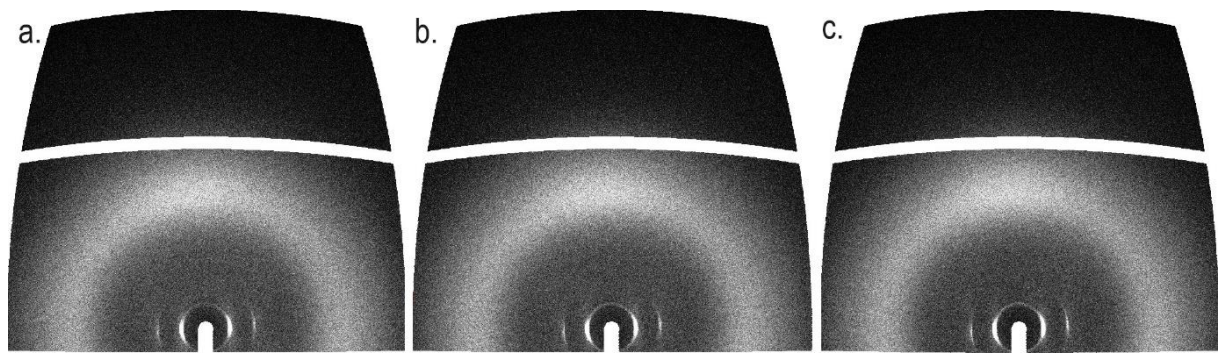

**Fig. S5:** X-ray scattering patterns of **3** in the (a) SmA, (b) SmA$_{AF}$ and (c) SmA$_{P2}$ phases.

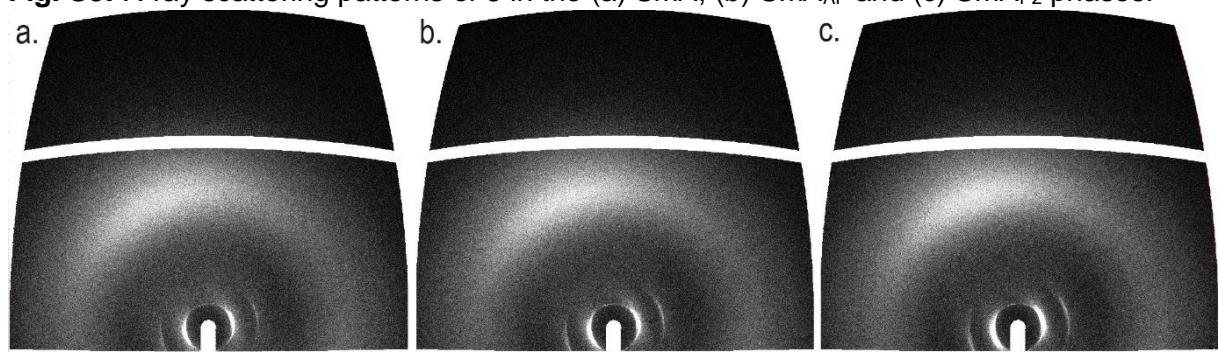

**Fig. S6:** Two dimensional X-ray scattering patterns of **4** in the (a) SmA, (b) SmA$_{AF}$ and (c) $SmC_P^H$ phases.

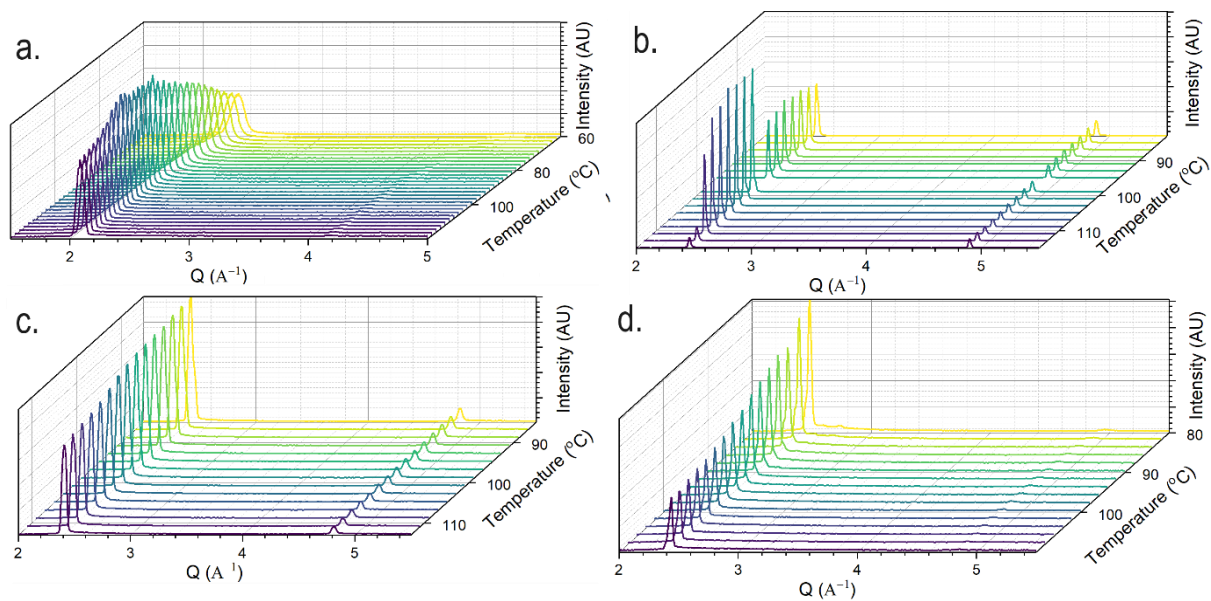

**Fig. S7:** 1D scattering patterns of (a) **1**, (b) **2**, (c), **3** and (d) **4** as a function of temperature.

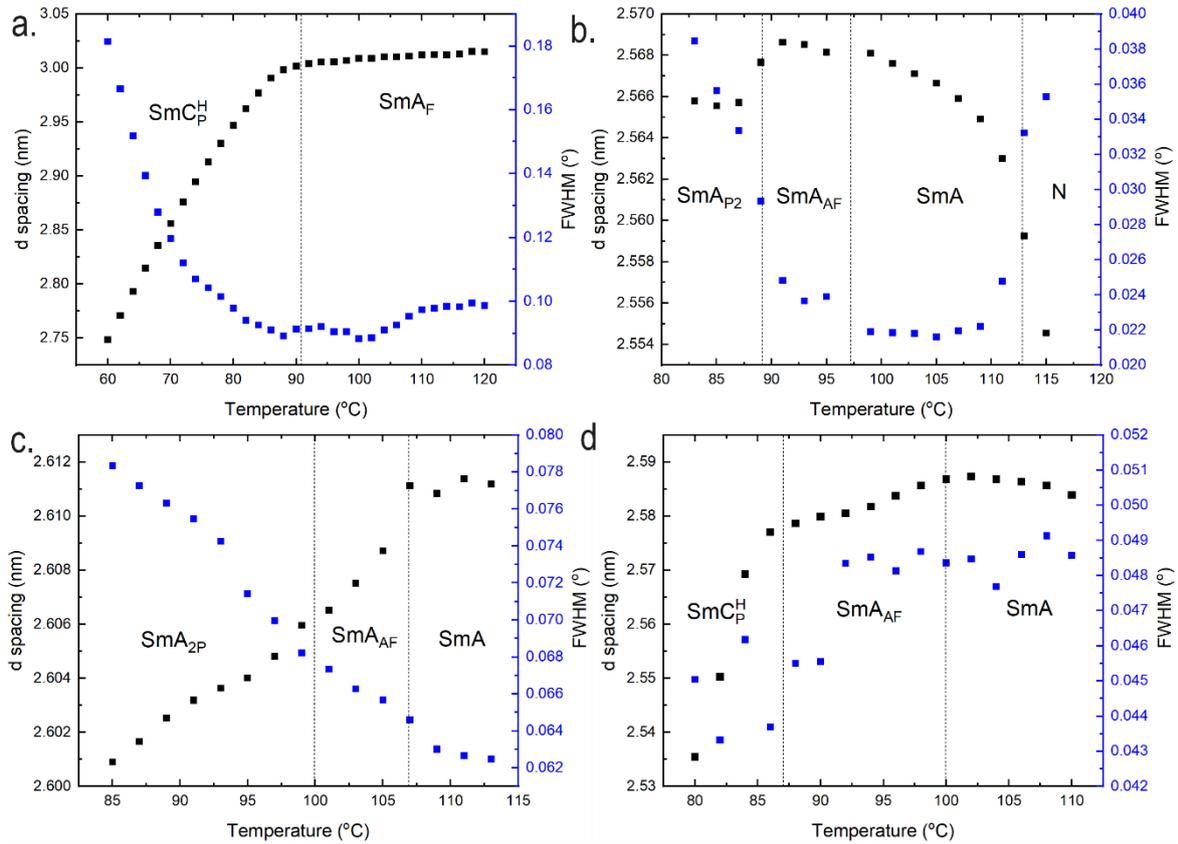

**Fig. S8:** *d* spacing and FWHM of the small angle peak from **Fig. S7** (above) of (a) **1**, (b) **2**, (c), **3** and (d) **4**.

The layer spacing for all the materials in phases we have designated as SmA type are comparable to the average molecular lengths obtained from DFT, indicating monolayer phase types. In the tilted phases (seen in materials **1** and **4**) the d spacing decreases from the DFT value indicating a tilted phase. All the materials show 2$^{nd}$ order Bragg scattering peaks for the small angle peak suggesting significant *pseudo* long range order in the layer spacing, consistent with what has been observed with the ferroelectric nematic phase [61].

## 2.3. Additional Spontaneous polarisation (P$_S$) Studies

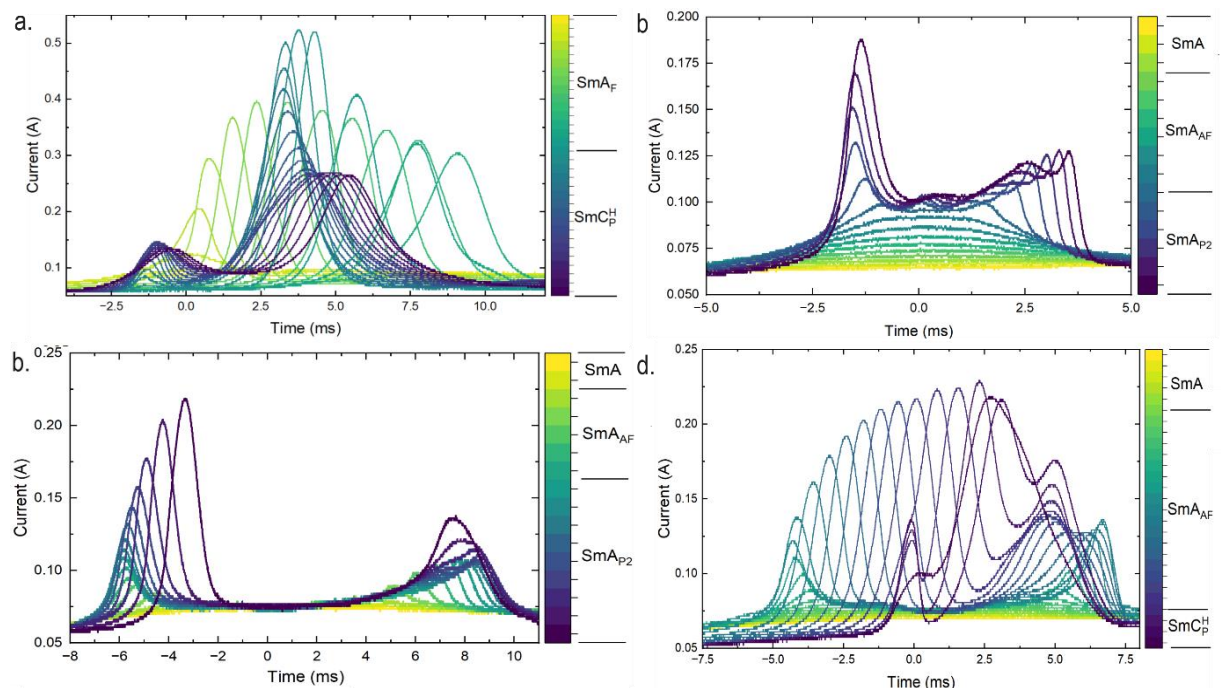

**Fig. S9**: Current response of the spontaneous polarisation (P$_s$) for (a) **1**, (b) **2**, (c) **3**, and (d) **4**. All measurements were performed at 20 Hz with the samples confined within a 4-micron cell with no alignment layer.

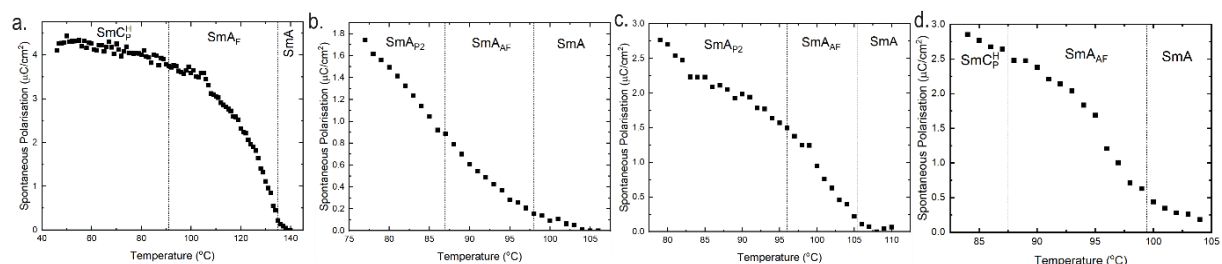

**Fig. S10**: Measured spontaneous polarisation (P$_s$) for (a) **1**, (b) **2**, (c) **3**, and (d) **4**. All measurements were performed at 20 Hz with the samples confined within a 4-micron cell with no alignment layer.

The four compounds studied here are show multiple polar smectic phases. Compound **1,** and to a lesser extent **4**, have measured P$_S$'s that are approaching saturation and as such have measured P$_S$ that gives a comparable result to the one obtained from MD simulations (**section S2.6**, **Table S2**). The P$_S$ values for compounds **2** and **3** are not saturated at the point of crystallisation and so are not comparable.

The SmA$_F$ phase was identified by the observation of only a single peak in the current response. Through the SmA$_F$ phase the time position of the peak continuously moved to longer timescales possibly due to increasing rotational viscosity and coercive voltage [62]. The SmA$_{AF}$ phase was identified by two peaks that are close to symmetric, with one occurring before the applied voltage polarity switch and the other after. The double peaks of the SmA$_{AF}$ phase also showed similar temperature dependence to the time position with the peak at negative times moving to increasingly negative times and the peak in the positive times moving to larger positive, i.e. increasing their separation.

The $\text{SmC}_\text{P}^\text{H}$ phase seemingly has a characteristic $P_S$ trace of a larger peak at longer times with a smaller peak that grows in at times even before the voltage has switched polarity (negative times). This peak is narrower than the longer timescale peak and continuously grows in area at the cost of the area of the larger peak. However, despite this population shift the total measured $P_S$ grows seemingly following the same temperature dependence being unaffected by this shift in relative area of the peaks. The peaks also shift in their time position both moving to more positive timescale with the smaller slower peak being able to shift to the positive times. We speculate that this peak could be associated with the tilt. Possibly only small amounts of applied voltage is required to remove the tilt which will induce a small amount of current flow. Upon voltage reversal the tilt could reform at low voltages again inducing a small amount of current flow. This peak grows continuously as the tilt increases (seen from the X-ray) due to the increased tilt angle allowing for larger changes in induced current flow.

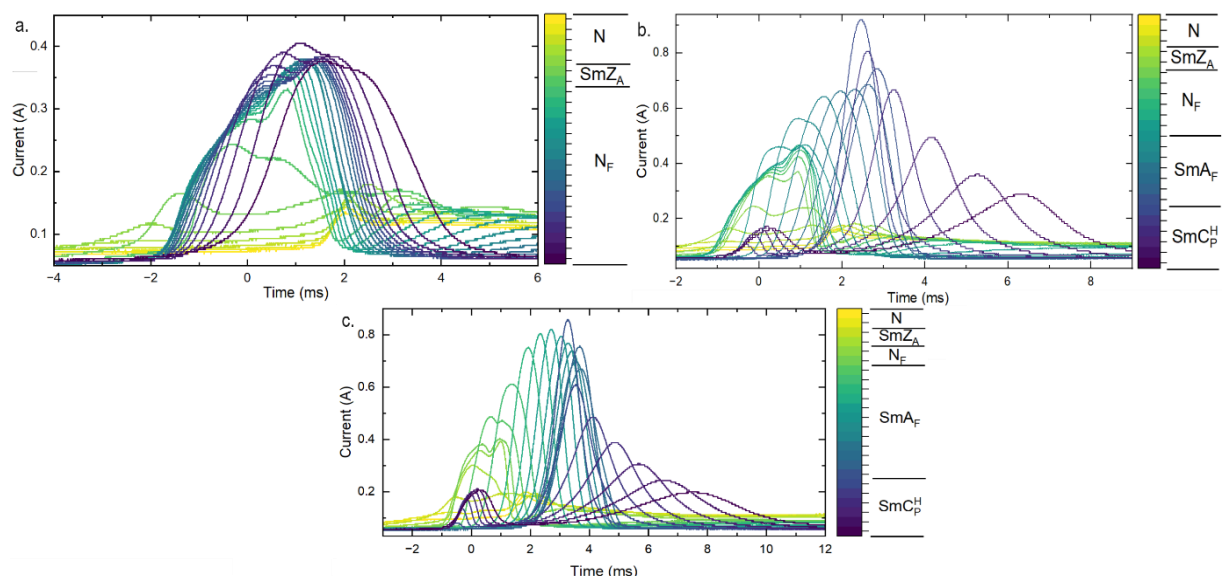

**Fig. S11**: Current response traces of **1** and **DIO** in molar ratios of: (a) 40:60, (b) 50:50 and (c) 60:40. All measurements were performed at 20 Hz with a voltage that saturated the measured $P_s$ and the samples confined within a 4-micron cell with no alignment layer.

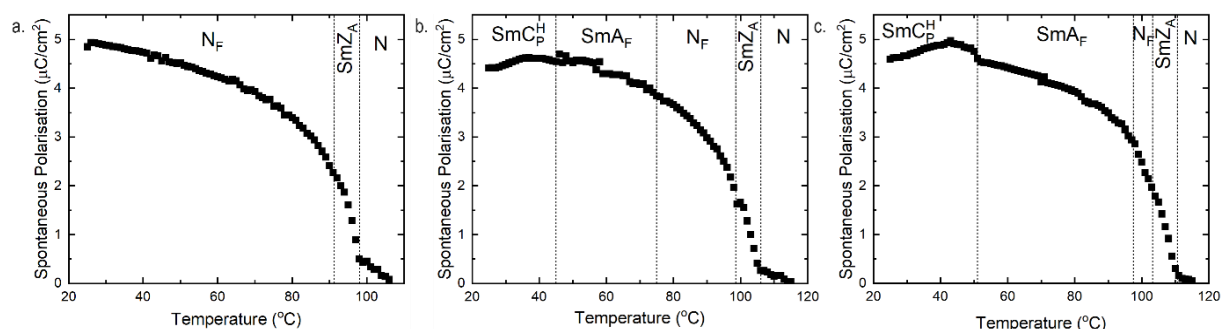

**Fig. S12**: Temperature dependence of spontaneous polarisation ($P_s$) of **1** and **DIO** in molar ratios of: (a) 40:60, (b) 50:50 and (c) 60:40. All measurements were performed at 20 Hz with the samples confined within a 4-micron cell with no alignment layer.

## 2.4. Additional Images of Selective Reflection

The selective reflection of light was not only observed in pure samples of **1** (**Fig. S13a**) but also in binary mixtures of **1** and **DIO** at ambient temperatures. As with the pure material, mixtures comprised of 60% **DIO** gave large domains of opposite handedness when viewed with circularly polarised light of opposite handedness (**Fig. S13b**). We were also able to obtain an image a thin cell with no anchoring condition containing a mixture comprised of 50% **DIO** showing almost the whole visible range at room temperature (**Fig. S13c**).

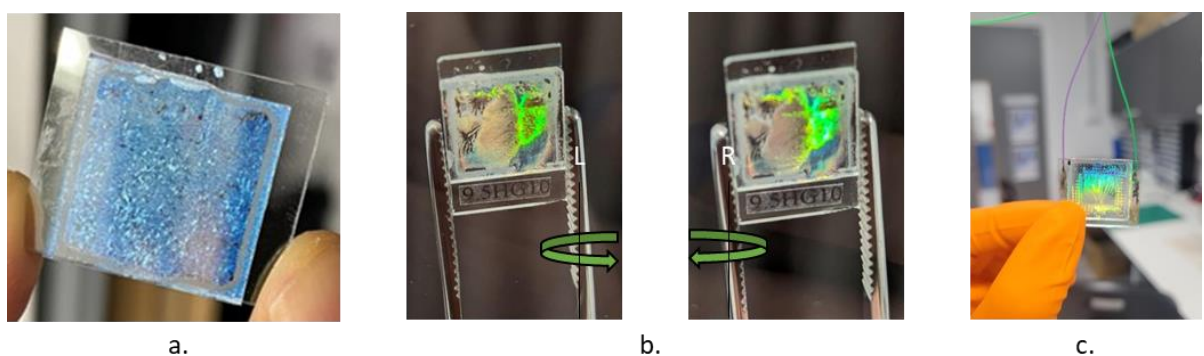

a.                    b.                    c.

**Fig. S13**: (a) Image of **1** confined in a 5-micron cell with no alignment layer showing the selective reflection of light and the many domains of the sample, (b) Images of binary mixtures of 60% **1** and **DIO** filled into a planar aligned cell taken with a circular polariser of left and right polarisation demonstrating the handedness of the light reflected from the sample. The lack of uniformity across the cell is indicative of various domains of different helical rotation and pitch orientation, and (c) a mixture of 50% **1** and **DIO** confined within a 10-micron cell with no alignment layer showing the selective reflection of light at room temperature.

## 2.5. Polarised Raman Spectroscopy

To gain greater insight into the ordering of **1** in the $\text{SmC}_\text{P}^\text{H}$ phase, Raman spectroscopy was performed to determine the 2$^{\text{nd}}$ and 4$^{\text{th}}$ rank uniaxial order parameters $\langle P_2 \rangle$ and $\langle P_4 \rangle$. Suggestions of heliconical tilt and/or tilt in general can be observed via a reduction in the determined order parameters $\langle P_2 \rangle_\text{app}$ and $\langle P_4 \rangle_\text{app}$ upon cooling into the tilted phase [39,63]. The cause of the reduction in $\langle P_2 \rangle_\text{app}$ and $\langle P_4 \rangle_\text{app}$ needs to be considered within the context of the observed phase. For example, for heliconical phases, it is expected that a reduction in $\langle P_2 \rangle_\text{app}$ and $\langle P_4 \rangle_\text{app}$ and no change in the director angle, $\theta_D$, as determined via fitting the depolarization ratio data [39,64] provided that the heliconical axis is in the same plane and direction as the original nematic director. It should be noted that for the N$_{\text{TB}}$ phase, close to the N to N$_{\text{TB}}$ transition, no reduction in $\langle P_2 \rangle$ was reported, most likely due to the very small tilt angle in the N$_{\text{TB}}$ phase, minimizing the effect, coupled with a likely small increase in the actual order parameter upon entering the lower temperature phase. However, a reduction in $\langle P_4 \rangle$ was observed and is related to the interplay between tilt and increasing orientational order on cooling [38].

For tilted phases, it is expected to see no reduction in $\langle P_2 \rangle_{\text{app}} = \langle P_2 \rangle$ and $\langle P_4 \rangle_{\text{app}} = \langle P_4 \rangle$ and no change in $\theta$ provided that the tilt happens within the plane of the laser polarization vector [65]. If, however, a component of the tilt occurs normal to the plane of the polarization vector, it is expected that a reduction of $\langle P_2 \rangle_{\text{app}}$ and $\langle P_4 \rangle_{\text{app}}$ would be observed. It should be noted that, in general, measurement of $\langle P_2 \rangle$ and $\langle P_4 \rangle$ can be determined in the tilted smectic phases, SmC* and de Vries SmA, but this requires special considerations such as determination of parameters like in plane tilt angle, a modification of the general Raman intensity functions, eq. **(1)** and eq. **(2)**, for uniaxial singly peaked ODF *ab initio*, or consideration of the alignment to suppress helical structure [66–68]. **Fig. S4a** shows the director angle as determined by the fitting of depolarization ratio data, it can be seen that the angle of the director remains effectively constant in any given domain. The tilt in the $\text{SmC}_P^H$ phase measured from SAXS layer spacing measurements, accounts for the reduction in the measured $\langle P_2 \rangle_{\text{app}}$ and $\langle P_4 \rangle_{\text{app}}$ as determined via PRS. As the director angle remains unchanged, it can be concluded that the phase structure is most likely a tilted director that is averaged around a symmetry axis e.g. in a helical manner within the phase.

Once $\langle P_2 \rangle_{\text{app}}$ and $\langle P_4 \rangle_{\text{app}}$ are determined, the heliconical tilt angle, $\Psi$, can be calculated via the following formula (**Fig. S14b**) [67,68]:

$$\langle P_n \rangle_{\text{app}} = \langle P_n \rangle P_n \cos(\Psi) \quad \textbf{(8)}$$

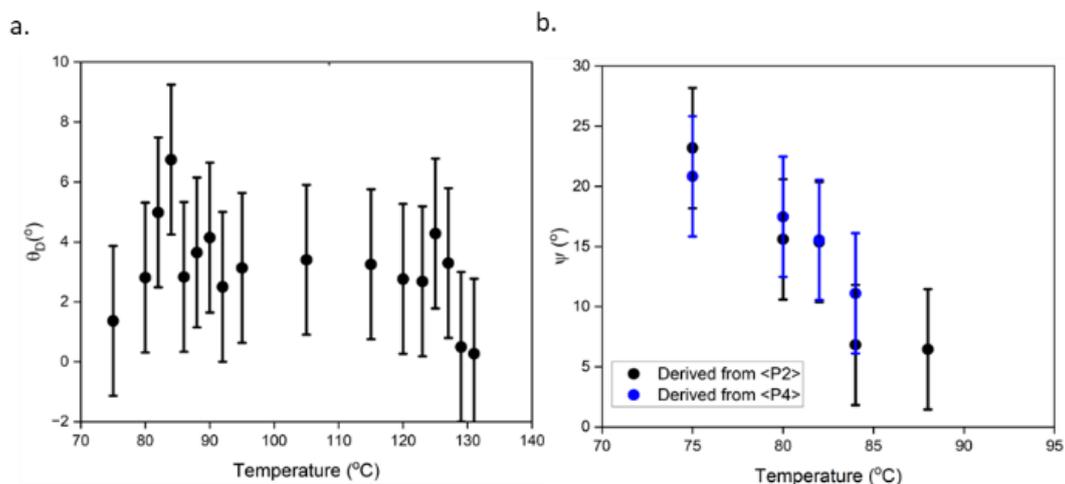

**Fig. S14**: (a) Director angle as determined via the fitting of the depolarization ratio eq. **(3)**, and (b) Heliconical angle as determined via PRS by fitting to eq. **(5)**.

## 2.6. DFT calculations and additional MD simulations

For **1**, the large molecular electric dipole moment (11.2 D) implies that the maximum polarisation possible would be ~ 5.5 µC cm$^2$ (assuming perfect polar order, $\langle P_1 \rangle = 1$, and a mass density of 1.3 g cm$^3$ [69] which is reasonably close to experimental values (**Table S1**). For **1** we also computed the Raman spectrum finding it to be in good agreement with the experimental spectrum (**Fig. S15a**) albeit slightly shifted to higher frequencies. Analysis of the displacement vectors confirms that the 1606 cm$^{-1}$ peak (used in the calculation of $\langle P_2 \rangle$, $\langle P_4 \rangle$) is associated with the breathing mode of the unfluorinated 1,4-disubstituted benzene (and thus is suitable for determination of $\langle P_2 \rangle$ and $\langle P_4 \rangle$), while the peak at 1630 cm$^{-1}$ coresponds to the breathing mode of the adjacent 3,5-difluorobenzene (**Fig. S15a**).

| Cmp. | No. Conf. | Dipole / D | Dipole Ang / ° | α / Å³ | Δα / Å³ |
|---|---|---|---|---|---|
| 1 | 46 | 11.2 | 9.0 | 62.6 | 50.8 |
| 2 | 33 | 9.5 | 12.0 | 54.5 | 46.9 |
| 3 | 37 | 9.2 | 10.0 | 55.0 | 48.9 |
| 4 | 29 | 8.5 | 13.0 | 54.7 | 46.4 |

**Table S1:** Tabulated properties of compounds **1-4** obtained from DFT calculations at the B3LYP-GD3BJ/aug-cc-pVTZ level of DFT given as the probability weighted average over the indicated number of conformers (No. Conf.). The dipole moment is the magnitude of the dipole tensor; Dipole Ang. Is the angle between the dipole moment and the molecular long axis, α and Δα are the isotropic and anisotropic polarizabilities, respectively.

Let us now discuss additional MD simulations of **1** as described in the manuscript. Each simulation commences from a polar nematic starting configuration, however compound **1** readily adopts a lamellar structure and so we observe a polar SmC phase at temperatures up to 430 K, while at 440 K and 450 K we observe a polar SmA phase (**Fig. S15c**). The transition temperatures are notably overestimated compared to experimental values, yet the ability to reproduce phase type is encouraging. Gratifyingly, the values of polarisation for these simulations are consistent with experimentally obtained values (and are within the limits estimated from DFT calculations), while the value of $\langle P_1 \rangle$ is close to saturated. The values of $\langle P_2 \rangle$ we obtain within each phase are consistent with the trends obtained experimentally by PRS. Depending on the temperature of the MD simulation, we find **1** to variously exhibit polar SmC, polar SmA, polar N, or isotropic phases. To facilitate comparison with experimental data (where this is available) we have offset the simulation temperature in the plots below so that they are relative to the $\mathrm{SmC_P^H}$ to SmA$_F$ transition (which occurs at 90 °C). It should be noted that our MD simulations overestimate this transition temperatures of **1** by around 50 °C. MD simulations competently predict the layer spacing in the SmA phase, as well as the temperature dependent decrease in the SmC phase (**Fig. S15c**). The polar order parameter $\langle P_1 \rangle$, remains close to saturated at all temperatures. If the orientational order parameter is computed relative to the layer normal, $\langle P_{2 layer normal} \rangle$, then it displays the same temperature dependent decrease as seen when measured *via* PRS. Additionally, we also performed MD simulations of compounds **2-4** at a temperature of 400 K (**Table S2**). Whereas **1** forms a polar SmC configuration spontaneously, we find **2** and **3** form a polar SmA phase (equivalent to the experimentally observed SmA$_F$ phase), whereas **4** remains in the polar nematic (equivalent to N$_F$) configuration. CDF plots were computed for **1-4** (**Fig. S15e**) at 400 K (variously in the polar SmC, SmA and N configurations). The CDF of **1** differs from **2-4**, with the head-to-tail separation being at a larger spacing owing to its increased molecular length (due to the additional 2,5-disubstituted 1,3-dioxane ring).

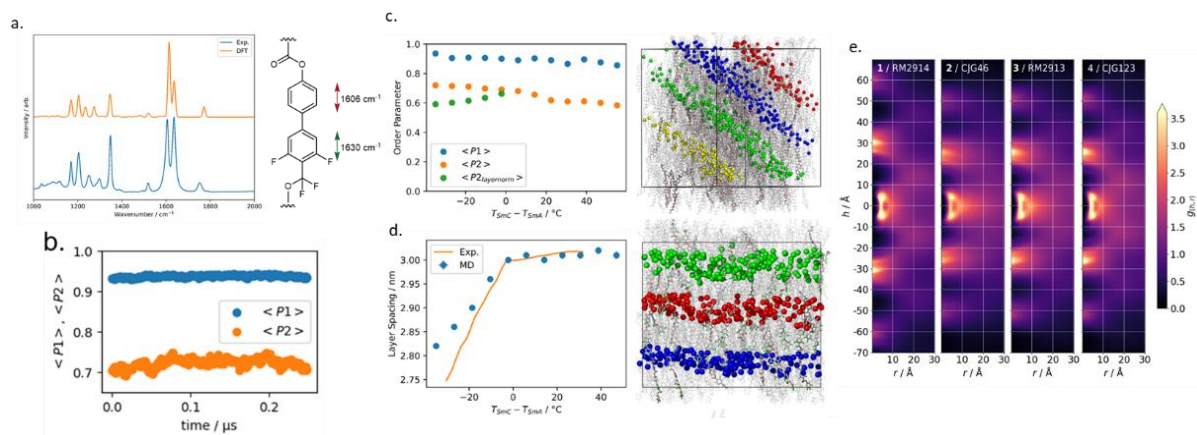

**Fig. S15:** (a) Plot of the experimental polarised Raman spectrum of **1** in the nematic phase (Exp., blue) and the simulated Raman spectrum of the same compound obtained at the B3LYP-GD3BJ/cc-pVTZ level of DFT (DFT, orange). The simulated spectrum is constructed by adding a Gaussian at each harmonic frequency whose height is the calculated Raman activity and with a FWHM of 15 cm$^{-1}$. The calculated and experimental spectra are offset for clarity, (b) lot of the polar (<P1>) and orientational (<P2>) order parameters as a function of simulation time, showing that both are sustained in the absence of the biasing field, (c+d) MD simulations for **1** showing plots of layer spacing and order parameters (polar, <P1>; orientational, <P2> and <P2$_{layernorm}$>) as a function of reduced temperature (defined as $T_{SmC} - T_{SmA}$) for (a) the polar SmC phase and (b) the polar SmA phase. The molecules are shown as wire frame, with the oxygen atom of the –CF$_2$O- group shown as a sphere to illustrate the layered structure and coloured according to the layer number, and (c) Cylindrical distribution functions (CDF) for **1-4** at 400 K In the polar SmC (**1**), SmA (**2, 3**) and N (**4**) configurations.

| Cmp. | Phase | <P1> | <P2> | Ps / C m$^2$ | ρ /kg m$^3$ | d / nm |
|---|---|---|---|---|---|---|
| 1 | SmC | 0.936 ± 0.003 | 0.718 ± 0.01 | 0.053 ± 0.0 | 1336.138 ± 2.452 | 2.8 ± 0.1 |
| 2 | SmA | 0.861 ± 0.006 | 0.647 ± 0.012 | 0.047 ± 0.0 | 1311.284 ± 3.612 | 2.5 ± 0.1 |
| 3 | SmA | 0.880 ± 0.005 | 0.670 ± 0.013 | 0.046 ± 0.0 | 1294.389 ± 3.254 | 2.4 ± 0.1 |
| 4 | N | 0.922 ± 0.004 | 0.692 ± 0.011 | 0.045 ± 0.0 | 1271.97 ± 3.387 | n/a |

**Table S2:** Tabulated simulation properties for compounds **1-4**, where *Phase* is the simulation phase type judged from inspection of the simulation, <P1> is the polar order parameter, <P2> is the orientational order parameter, P$_S$ is the spontaneous polarization, ρ is the density, and d is the layer spacing. All simulations were performed at 400 K and commenced from a polar nematic starting configuration, and in all cases polar order was retained over the entire production MD simulation (250 ns).

### 2.7. The SmA$_{P2}$ phases

The focus of this paper is on introducing the $SmC_P^H$ phase as well as the SmA$_{AF}$ phase. We do however observe two other polar SmA phases which we simply designate as the SmA$_{P2}$ phase from their presence as the 2$^{nd}$ polar SmA phase for those materials, specifically **2** and **3**. Although there is little textural variation between the SmA$_{P2}$ and the SmA$_{AF}$ phase, making it had to notice via POM, there is a weak thermal transition observable via DSC and slight changes in the SAXS d-spacing indicating changes in the molecular packing. More noticeable is the changes in the current response to an applied triangular wave voltage. For both materials the symmetric anti-ferroelectric peaks in the SmA$_{AF}$ phase preceding the SmA$_{P2}$ phase become non-symmetric in the lower phase with the peak at negative times having the larger associated area. This change could indicate a switch to a ferri-electric phase but could equally be explained by some other change effecting the switching behaviour of the materials. Thorough investigation of these materials is a most attractive prospect for future study.

# 3 Organic Synthesis

Materials **1-4** were synthesised as outlined in **Scheme S1.** A Suzuki-Miyaura cross-coupling of 5-bromo-2-(difluoro(3,4,5-trifluorophenoxy)methyl)-1,3-difluorobenzene with 4-hydroxybenzene boronic acid pinacol ester afforded 5-(4-hydroxyphenyl)-2-(difluoro(3,4,5-trifluorophenoxy)methyl)-1,3-difluorobenzene (*i1*) in high yield. Subsequent esterification with a selection of carboxylic acids afforded compounds **1 – 4**.

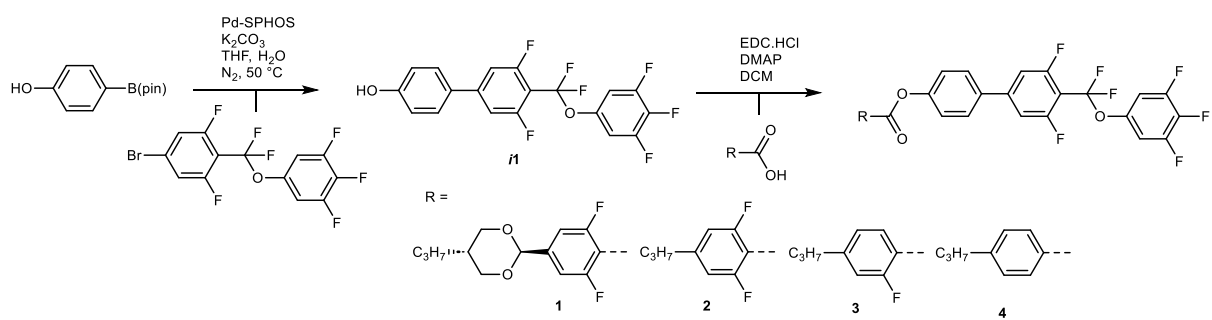

**Scheme S1**: Synthetic scheme for the preparation of materials **1-4.**

## 3.1 Synthesis of 5-(4-hydroxyphenyl)-2-(difluoro(3,4,5-trifluorophenoxy)methyl)-1,3-difluorobenzene (*i1*)

A solution of 5-bromo-2-(difluoro(3,4,5-trifluorophenoxy)methyl)-1,3-difluorobenzene (7.78 g, 20 mmol) in a biphasic mixture of THF (60 mL) and 2M aqueous K$_2$CO$_3$ (60 mL, 2M) was degassed by sparging with argon. Separately, 10 mL of THF was degassed by sparging with argon; solid Pd(OAc)$_2$ (1 mmol) and SPHOS (2 mmol) were added, the suspension was stirred for 5 minutes to afford a solution of Pd-SPHOS in THF. The biphasic reaction solution was heated under reflux under an atmosphere of dry argon; 4-hydroxyphenyl boronic acid pinacol ester (4.84 g, 22 mmol) was added in one portion, followed by the solution of Pd-SPHOS in THF. The reaction was heated under reflux for 12 h, at which point TLC showed complete consumption of the starting pinacol ester. The solution was then cooled to ambient temperature, the aqueous layer separated and washed with ethyl acetate (3x 50 ml) and discarded. The combined organics were sequentially washed with saturated aqueous ammonium carbonate (50 ml), brine (50 ml). The organics were then dried over MgSO$_4$, filtered, and volatiles removed in vacuo. The crude material was filtered over a short plug of silica gel, eluting with DCM, and then recrystalised from ethanol, affording the title compound.

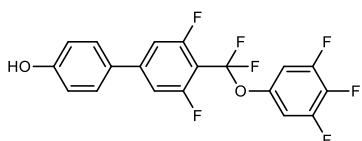

5-(4-hydroxyphenyl)-2-(difluoro(3,4,5-trifluorophenoxy)methyl)-1,3-difluorobenzene (*i1*)

Yield: 7.2 g (89 %, off white solid);

R$_f$ (DCM): 0.33

$^1$H NMR (400 MHz, CDCl$_3$): 5.70 (1H, S, Ar-O**H**), 6.96 (2H, ddd, $J_{H-H}$ = 2.4 Hz, $J_{H-H}$ = 2.8 Hz, $J_{H-H}$ = 8.4 Hz, Ar-**H**), 6.97 - 7.02 (2H, m, Ar-**H**), 7.15-7.21 (2H, m, Ar-**H**), 7.49 (2H, ddd, $J_{H-H}$ = 2.4 Hz, $J_{H-H}$ = 2.8 Hz, $J_{H-H}$ = 8.4 Hz, Ar-**H**).

$^{19}$F NMR (376 MHz, CDCl$_3$): -163.27 (1F, tt, $J_{H-F}$ = 6.1 Hz, $J_{F-F}$ = 21.0 Hz, Ar-**F**), -132.54 (2F, dd, $J_{H-F}$ = 8.7 Hz, $J_{F-F}$ = 20.5 Hz, Ar-**F**), -110.71 (2F, td, $J_{H-F}$ = 11.5 Hz, $J_{F-F}$ = 26.2 Hz, Ar-**F**), -61.51 (2H, t, $J_{F-F}$ = 26.2 Hz, C**F$_2$**O).z

### 3.2 General esterification procedure used in the synthesis of 1-4

A round bottomed flask or 14ml vial was charged with 5-(4-hydroxyphenyl)-2-(difluoro(3,4,5-trifluorophenoxy)methyl)-1,3-difluorobenzene (*i1*) (402 mg, 1 mmol), the appropriate carboxylic acid (1.1 mmol), EDC.HCl (1.5 mmol) and DMAP (0.1 mmol). Dichloromethane was added (conc. ~ 0.1 M) and the suspension stirred until complete consumption of the phenol as judged by TLC. The volatiles were removed in vacuo and the crude material was subjected to flash chromatography over silica gel with a gradient of hexane/DCM using a Combiflash NextGen 300+ System (Teledyne Isco). The materials were subsequently recrystallized from EtOH before being dried under reduced pressure to give the reported yields.

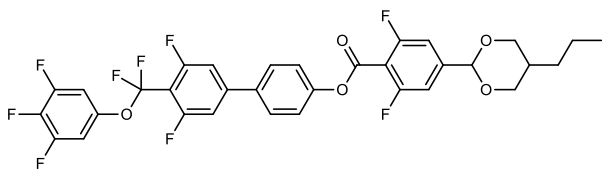

**(1)**

Yield: 490 mg (74 %, colourless crystals);

R$_f$ (DCM): 0.90

$^1$H NMR (500 MHz, CDCl$_3$): δ 7.62 (ddd, $J$ = 8.7, 2.6, 1.9 Hz, 2H, Ar-**H**), 7.39 (ddd, $J$ = 8.6, 2.7, 2.0 Hz, 2H, Ar-**H**), 7.24 – 7.16 (m, 4H. Ar-**H**), 7.04 – 6.94 (m, 2H, Ar-**H**), 5.40 (s, 1H, Ar-C**H**-(O)$_2$), 4.26 (dd, $J$ = 7.0, 4.6 Hz, 2H, O-C**H**$_{axial}$(H)-CH), 3.54 (t, $J$ = 11.4 Hz, 2H, O-C**H**$_{eqit}$(H)-CH), 2.15 (m, 1H, (CH$_2$)$_2$-C**H**-CH$_2$), 1.35 (h, $J$ = 7.2 Hz, 2H, CH$_2$-C**H$_2$**-CH$_3$), 1.11 (q, $J$ = 7.2 Hz, 2H, CH-C**H$_2$**-CH$_2$), 0.94 (t, $J$ = 7.3 Hz, 3H, CH$_2$-C**H$_3$**).

$^{13}$C{$^1$H} NMR (101 MHz, CDCl$_3$):162.20, 162.15, 161.56, 159.64, 159.61, 159.58, 159.04, 152.29, 152.19, 151.27, 149.85, 149.80, 149.74, 149.69, 146.10, 145.99, 145.89, 145.54, 145.45, 145.35, 144.67, 139.71, 137.37, 137.22, 137.06, 135.55, 128.24, 122.90, 122.50, 120.25, 117.61, 111.20, 111.17, 110.96, 110.93, 110.40, 110.36, 110.16, 110.13, 109.93, 109.76, 109.58, 108.42, 107.58, 107.51, 107.40, 107.34, 98.84, 77.35, 77.03, 76.72, 72.59, 33.90, 30.23, 19.53, 14.18.

$^{19}$F NMR (376 MHz, CDCl$_3$): -61.66 (t, $J$ = 26.1 Hz, C**F$_2$**O), -108.69 (d, $J$ = 9.8 Hz, Ar-**F**), -110.06 (td, $J$ = 26.2, 11.0 Hz, Ar-**F**), -132.46 (dd, $J$ = 20.8, 8.7 Hz, Ar-**F**), -163.16 (tt, $J$ = 21.1, 5.9 Hz, Ar-**F**).

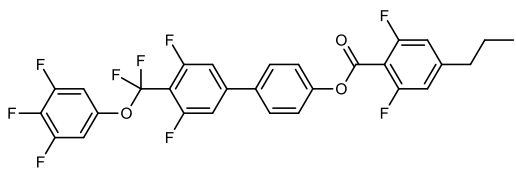

**(2)**

Yield: 414 mg (71 %, white crystals);

R$_f$ (DCM): 0.80

$^1$H NMR (400 MHz, CDCl$_3$): 7.62 (d, *J* = 8.8 Hz, 2H, Ar-**H**), 7.38 (d, *J* = 8.7 Hz, 2H, Ar-**H**), 7.22 (d, *J* = 10.3 Hz, 2H, Ar-**H**), 7.04 – 6.94 (m, 2H, Ar-**H**), 6.87 (d, *J* = 9.3 Hz, 2H, Ar-**H**), 2.65 (t, *J* = 7.3 Hz, 2H, Ar-C**H$_2$**-CH$_2$), 1.68 (h, *J* = 7.5 Hz, 2H, CH$_2$-C**H$_2$**-CH$_3$), 0.97 (t, *J* = 7.4 Hz, 3H, CH$_2$-C**H$_3$**).

$^{13}$C{$^1$H}{$^{19}$F} NMR (128 MHz, CDCl$_3$: 162.21, 162.16, 161.29, 160.15, 160.11, 159.92, 159.24, 152.02, 151.34, 150.92, 150.84, 146.11, 146.02, 137.47, 135.46, 128.31, 128.23, 122.66, 122.55, 122.40, 120.25, 112.35, 112.32, 112.18, 112.15, 111.16, 110.96, 107.56, 107.51, 107.42, 107.37, 107.06, 77.29, 77.03, 76.78, 37.87, 23.66, 13.56.

$^{19}$F NMR (376 MHz, CDCl$_3$): -61.66 (t, *J* = 26.2 Hz, C**F$_2$**O), -109.64 (d, *J* = 10.1 Hz, Ar-**F**), -110.07 (td, *J* = 26.3, 11.3 Hz, Ar-**F**), -132.46 (dd, *J* = 20.6, 8.7 Hz, Ar-**F**), -163.14 (tt, *J* = 20.7, 6.0 Hz, Ar-**F**).

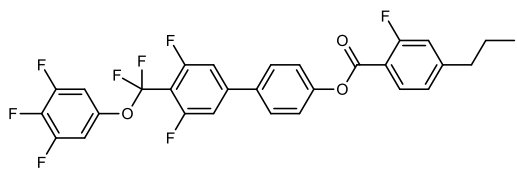

**(3)**

Yield: 340 mg (62 %, colourless crystals);

R$_f$ (DCM): 0.95

$^1$H NMR (400 MHz, CDCl$_3$) δ 7.95 (t, *J* = 7.8 Hz, 1H, Ar-**H**), 7.54 (ddd, *J* = 8.8, 2.7, 2.1 Hz, 2H, Ar-**H**), 7.28 (ddd, *J* = 8.8, 2.7, 2.2 Hz, 2H, Ar-**H**), 7.17 – 7.09 (m, 2H, Ar-**H**), 7.05 – 7.00 (m, 1H, Ar-**H**), 7.00 – 6.96 (m, 1H, Ar-**H**), 6.97 – 6.87 (m, 3H, Ar-**H**), 2.60 (tz, *J* = 8.5, 6.8 Hz, 2H, Ar-C**H$_2$**-CH$_2$), 1.62 (h, *J* = 7.5 Hz, 2H, CH$_2$-C**H$_2$**-CH$_3$), 0.90 (t, *J* = 7.4 Hz, 3H, CH$_2$-C**H$_3$**).

$^{13}$C{$^1$H} NMR (101 MHz, CDCl$_3$): 163.79, 162.64, 162.60, 161.58, 161.19, 159.00, 152.14, 152.05, 151.67, 149.85, 146.23, 146.12, 146.02, 144.68, 139.71, 137.22, 135.18, 132.41, 128.16, 124.48, 124.45, 122.66, 120.26, 117.15, 116.93, 114.96, 114.87, 111.16, 111.13, 110.92, 110.89, 107.57, 107.51, 107.34, 77.34, 77.23, 77.03, 76.71, 37.85, 23.89, 13.66.

$^{19}$F NMR (376 MHz, CDCl$_3$): -61.64 (t, *J* = 26.3 Hz, C**F$_2$**O), -108.49 (dd, *J* = 11.9, 7.6 Hz, Ar-**H**), -110.13 (td, *J* = 26.3, 11.0 Hz, Ar-**H**), -132.46 (dd, *J* = 20.6, 8.6 Hz, Ar-**H**), -163.15 (tt, *J* = 20.7, 5.9 Hz, Ar-**H**).

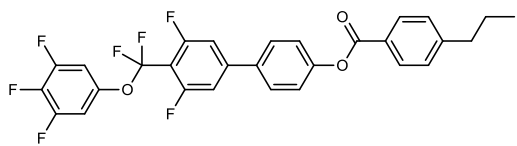

**(4)**

Yield: 470 mg (86 %, white needles);

R$_f$ (DCM): 0.83

1H NMR (400 MHz, CDCl3) δ 8.13 (ddd, *J* = 8.2, 1.7, 1.5 Hz, 2H, Ar-**H**), 7.63 (ddd, *J* = 8.7, 2.9, 1.9 Hz, 2H, Ar-**H**), 7.39 – 7.30 (m, 4H, Ar-**H**), 7.22 (m, 2H, Ar-**H**), 7.05 – 6.95 (m, 2H, Ar-**H**), 2.70 (t, *J* = 6.8 Hz, 2H, Ar-C**H$_2$**-CH$_2$),), 1.70 (h, *J* = 7.2 Hz, 2H, CH$_2$-C**H$_2$**-CH$_3$), 0.98 (t, *J* = 7.3 Hz, 3H, CH$_2$-C**H$_3$**).

$^{13}$C{$^1$H} NMR (101 MHz, CDCl$_3$): 165.07, 161.61, 159.05, 152.35, 152.30, 152.19, 152.05, 149.86, 149.80, 149.49, 146.27, 146.17, 146.07, 144.68, 139.71, 137.22, 135.03, 130.33,

128.83, 128.17, 126.63, 122.70, 120.27, 117.63, 111.14, 111.11, 110.90, 110.87, 107.58, 107.51, 107.34, 38.15, 24.26, 13.75.
$^{19}$F NMR (376 MHz, CDCl$_3$): -61.64 (t, *J* = 26.3 Hz, C**F$_2$**O), -110.14 (td, *J* = 26.2, 11.1 Hz, Ar-**H**), -132.45 (dd, *J* = 20.6, 8.7 Hz, Ar-**H**), -163.15 (tt, *J* = 21.0, 5.9 Hz, Ar-**H**).

## 3.3 Example NMR Spectra

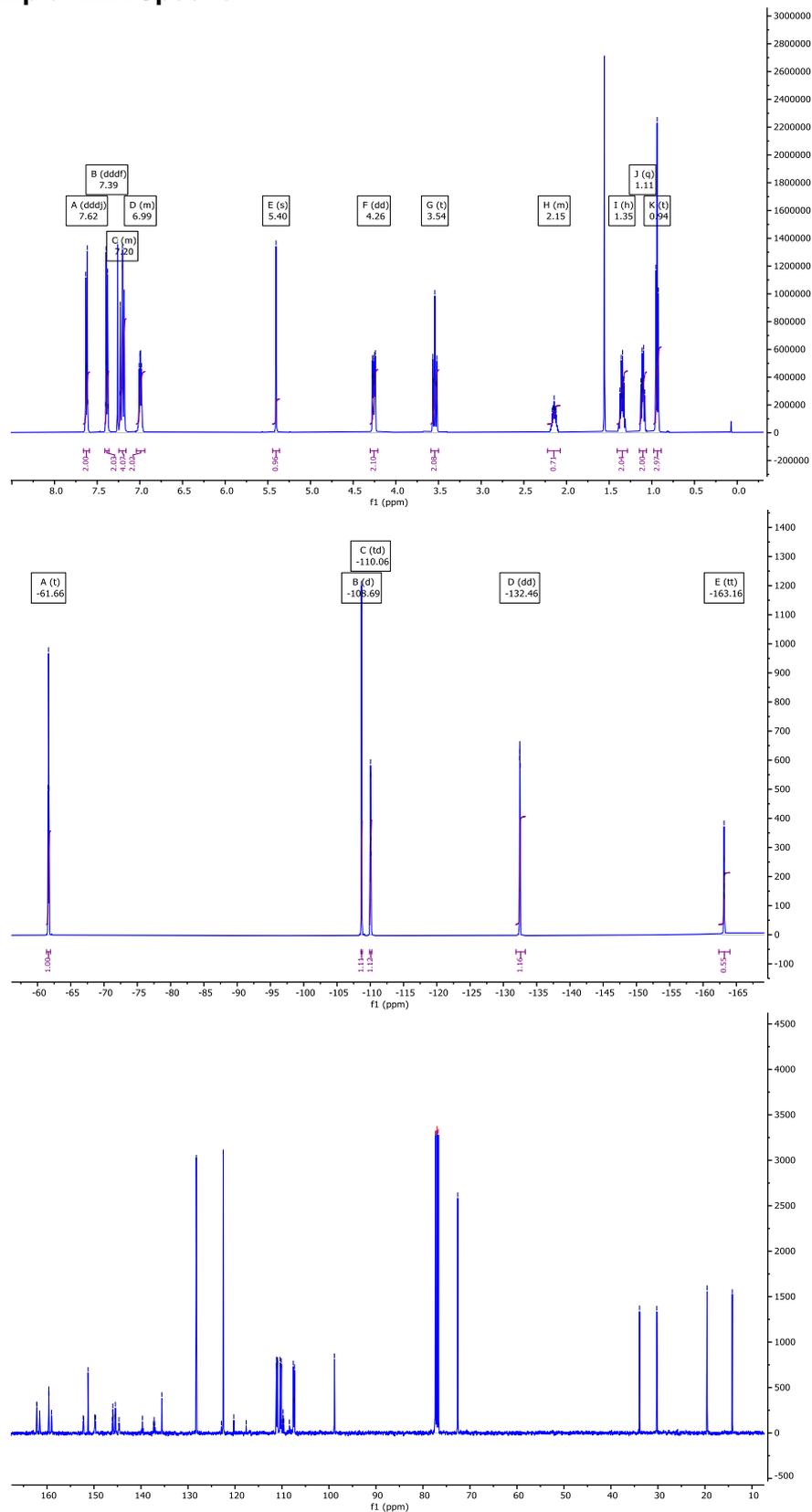

**Fig. S16**: Example NMR spectra of material **1**: $^1$H [top], $^{19}$F [middle], and $^{13}$C{$^1$H} [bottom].

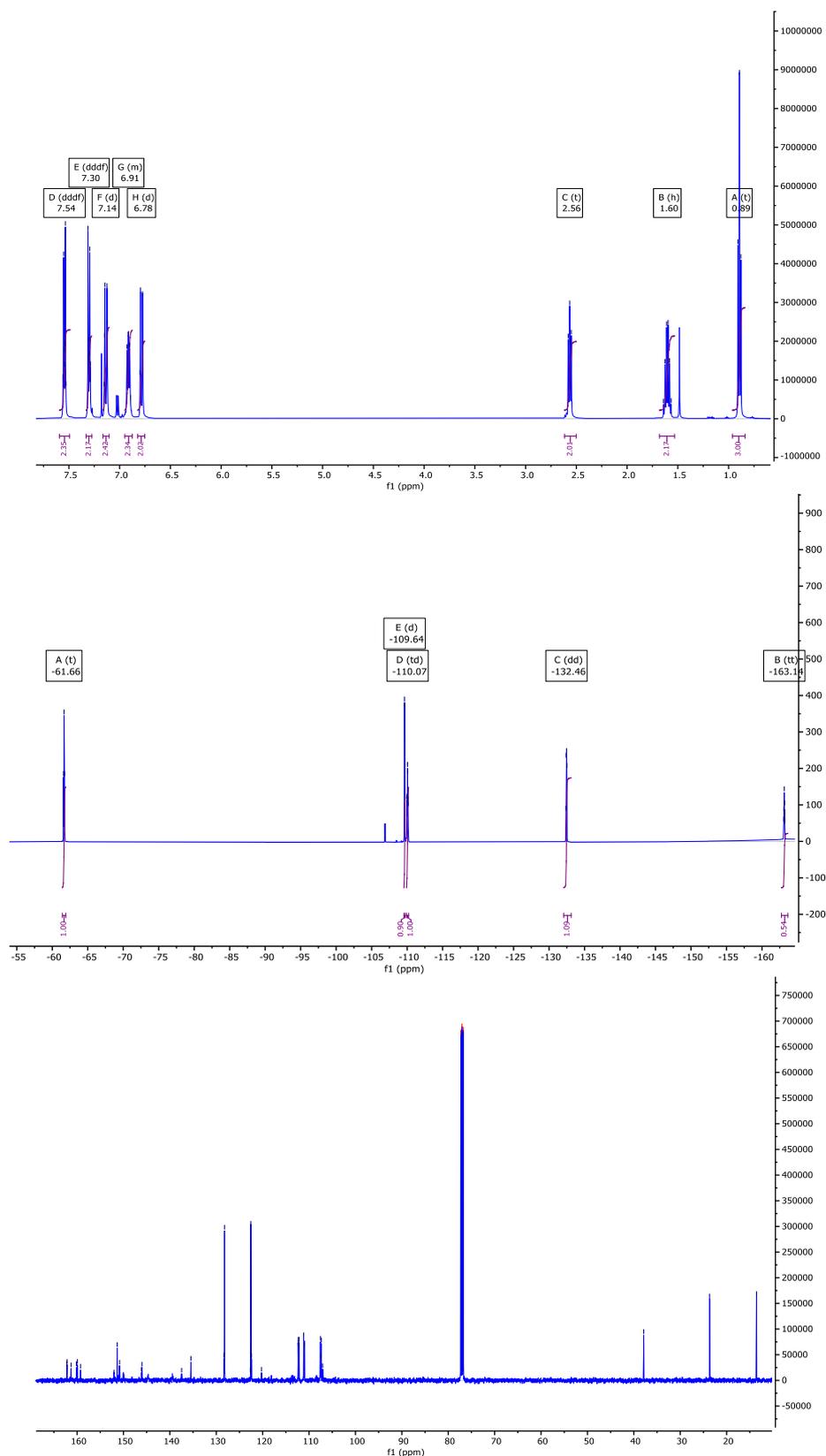

**Fig. S17**: Example NMR spectra of material **2**: $^1$H [top], $^{19}$F [middle], and $^{13}$C{$^1$H} [bottom].

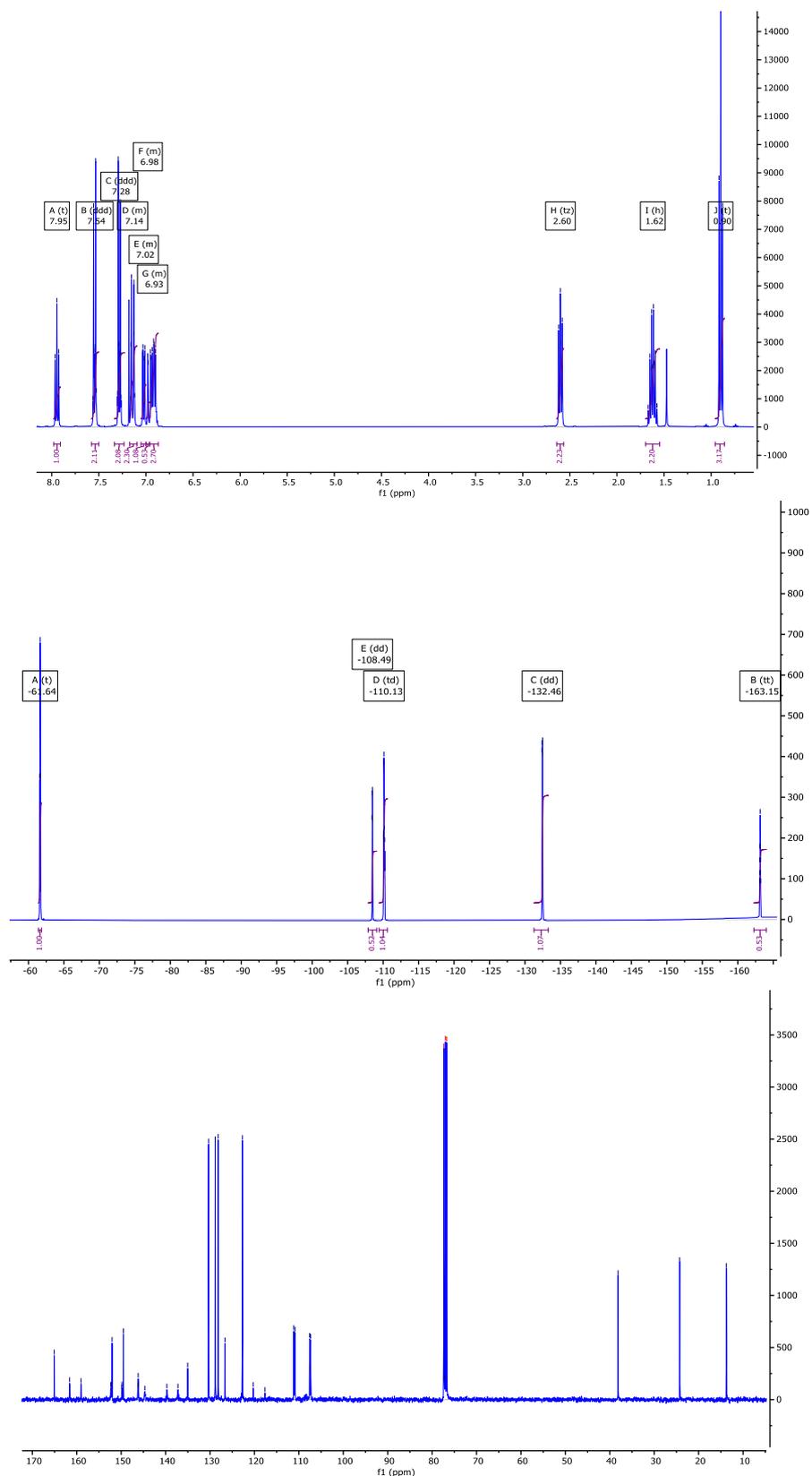

**Fig. S18**: Example NMR spectra of material **3**: $^1$H [top], $^{19}$F [middle], and $^{13}$C{$^1$H} [bottom].

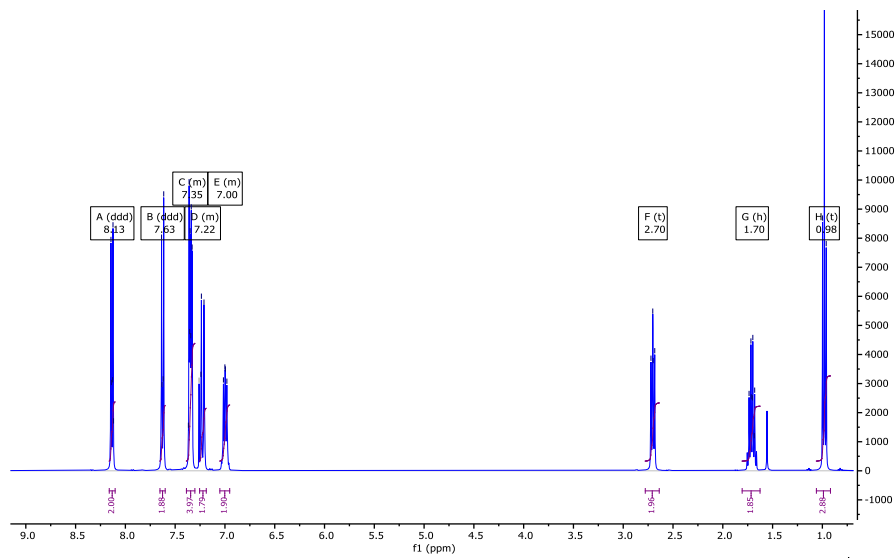
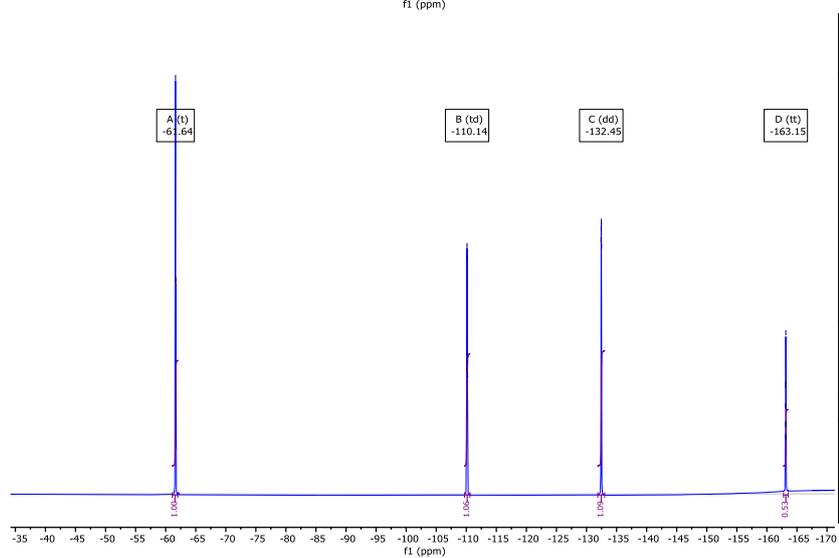
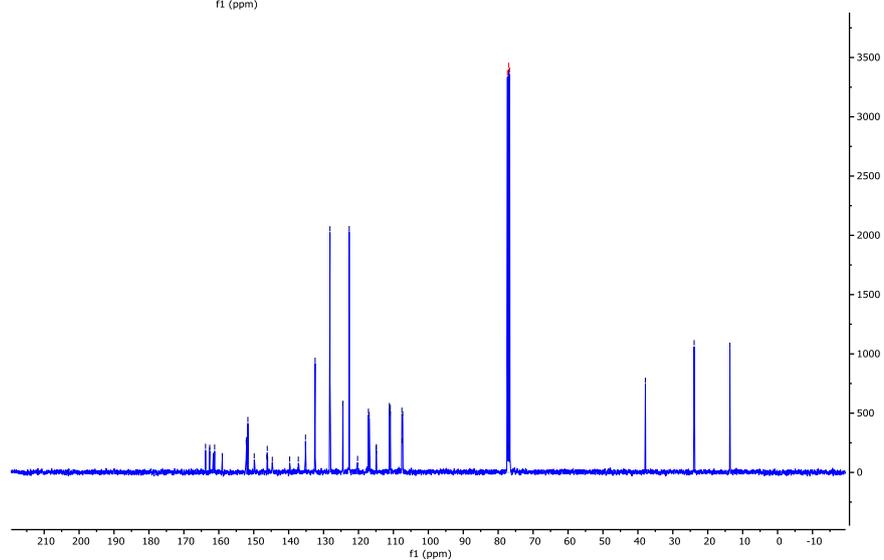

**Fig. S19**: Example NMR spectra of material **4**: $^1$H [top], $^{19}$F [middle], and $^{13}$C{$^1$H} [bottom].